\documentclass{egpubl}
\usepackage{egsgp2023}
 
\ConferencePaper        
\SpecialIssuePaper         

\CGFccby

\usepackage[T1]{fontenc}
\usepackage{dfadobe}  

\biberVersion
\BibtexOrBiblatex
\usepackage[backend=biber,bibstyle=EG,citestyle=alphabetic,backref=true]{biblatex} 
\electronicVersion
\PrintedOrElectronic

\ifpdf \usepackage[pdftex]{graphicx} \pdfcompresslevel=9
\else \usepackage[dvips]{graphicx} \fi

\usepackage{egweblnk} 

\usepackage{amsmath}
\usepackage{amssymb}

\usepackage{color}
\usepackage{url}
\usepackage[ruled]{algorithm2e} 

\usepackage{xcolor}

\newcommand{\MB}[1]{}

\newcommand{\rev}[1]{{\color{black} #1}}

\newcommand{\xR}{\mathbb{R}}
\newcommand{\xC}{\mathbb{C}}
\newcommand{\xCe}{\hat\xC}
\newcommand{\conj}[1]{\bar{#1}}
\newcommand{\mob}[4]{\frac{#1 z + #2}{#3 z + #4}}
\newcommand{\tin}{\!\in\!}

\newcommand{\md}{\delta} 
\newcommand{\lmd}{\ell} 
\newcommand{\bc}[1]{B_{#1}} 

\newcommand{\mobius}{M{\"o}\-bi\-us\xspace}
\newcommand{\element}[1]{\mathcal{#1}}

\newcommand{\coord}[1]{#1}
\newcommand{\dmap}[1]{#1}
\newcommand{\cmap}[1]{\overline{#1}}

\newcommand{\mM}{\element{M}}
\newcommand{\mV}{\element{V}}
\newcommand{\mE}{\element{E}}
\newcommand{\mT}{\element{T}}

\title{BPM: Blended Piecewise M{\"o}\-bi\-us\xspace Maps}

\author[Shir Rorberg \& Amir Vaxman  \& Mirela Ben-Chen]
{\parbox{\textwidth}{\centering Shir Rorberg$^{1}$\orcid{0000-0003-1080-8487}
        Amir Vaxman$^{2}$\orcid{0000-0001-6998-6689} 
        Mirela Ben-Chen$^{1}$\orcid{0000-0002-1732-2327} 
        }
        \\
{\parbox{\textwidth}{\centering $^1$Technion - Israel Institute of Technology\\
         $^2$The University of Edinburgh
       }
}
}

\begin{document}

\teaser{
\includegraphics[width=.94\textwidth]{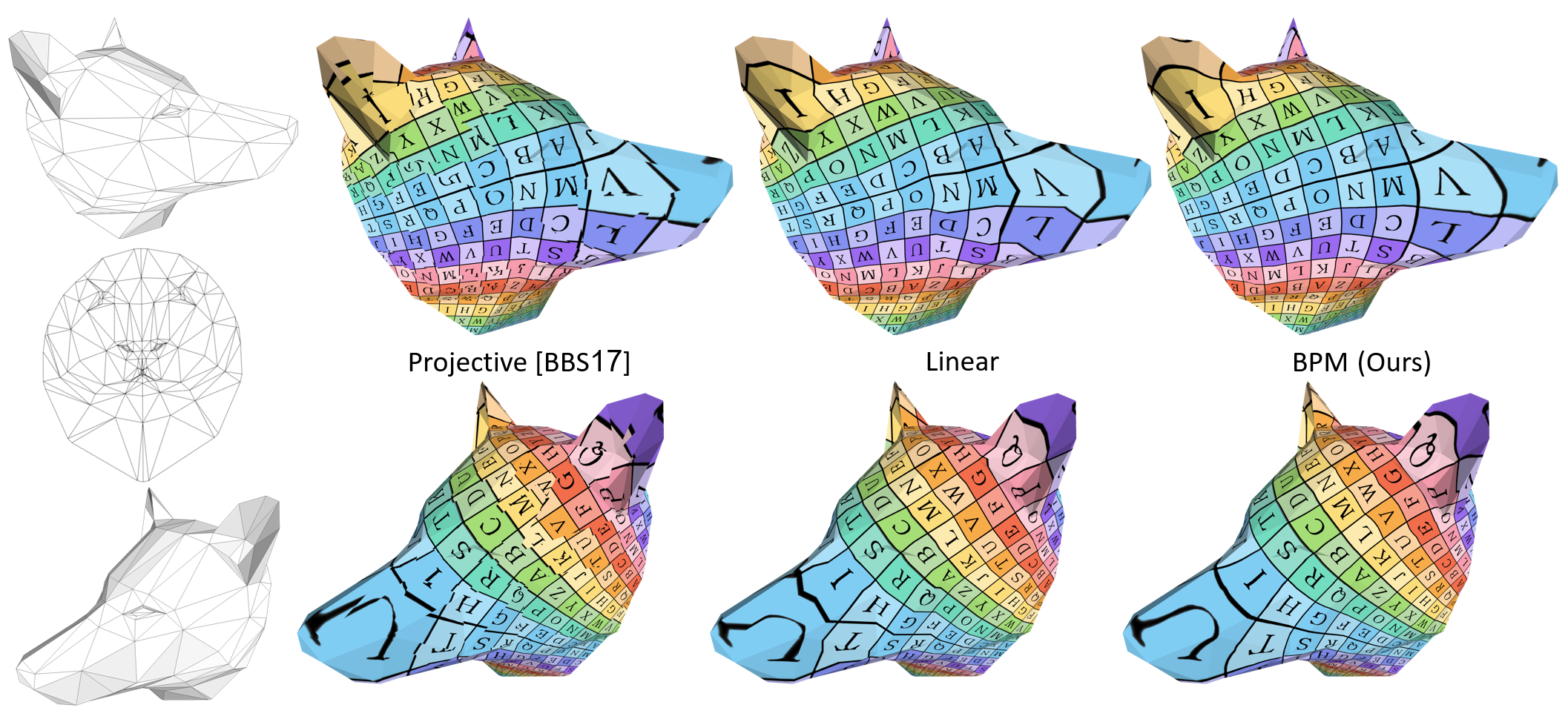} \centering
  \caption{  
  \rev{We propose an interpolation method that generates a continuous map between a mesh and its discrete planar parameterization (left).} Even for coarse meshes, our result (right) produces a smooth map which is superior to the default linear interpolation (middle) and the projective interpolation by~\cite{born2017quasiconformal} (left).} 
\label{fig:teaser}
}


\maketitle
\begin{abstract}
We propose a novel \emph{\mobius interpolator} that takes as an input a discrete map between the vertices of two planar triangle meshes, and outputs a \emph{continuous} map on the input domain. The output map \emph{interpolates} the discrete map, is continuous between triangles, and has low quasi-conformal distortion when the input map is discrete conformal.
Our map leads to considerably smoother texture transfer compared to the alternatives, even on very coarse triangulations. Furthermore, our approach has a closed-form expression, is local, applicable to \emph{any} discrete map, and leads to smooth results even for extreme deformations. Finally, by working with local intrinsic coordinates, our approach is easily generalizable to discrete maps between a surface triangle mesh and a planar mesh, i.e., a planar parameterization. We compare our method with existing approaches, and demonstrate better texture transfer results, and lower quasi-conformal errors.

\end{abstract}  

\section{Introduction}
Given two triangle meshes with the same connectivity, a natural vertex-to-vertex map is induced by the shared connectivity. In addition, a natural triangle-to-triangle map is induced by the unique linear map between corresponding triangles. These \emph{piecewise linear} maps are used almost exclusively in graphics and geometry applications to transfer quantities such as texture between meshes with the same connectivity. 

\begin{figure}[b]
    \centering
    \includegraphics[width=1.0\linewidth]{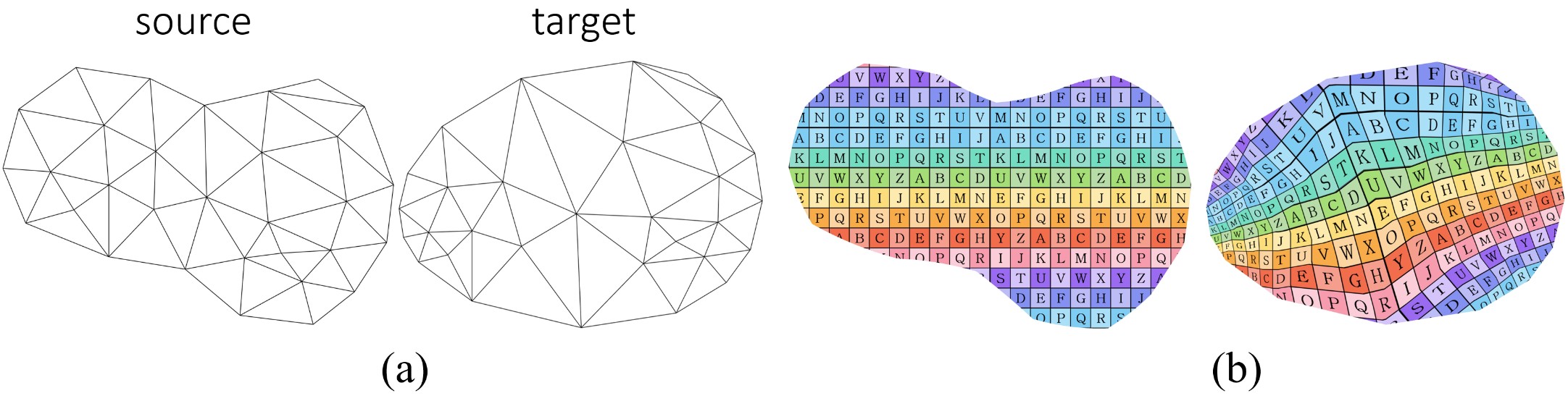}
    \caption{Piecewise-Linear map of a CETM vertex-to-vertex map \rev{\cite{springborn2008conformal}}. The input vertex to vertex map (a) and the pullback of the texture (b). Note the large angular distortion.} 
    \label{fig:pl_is_bad}
\end{figure}

While simple, piecewise linear maps lead to visible discontinuities when applied to coarse triangulations that undergo large deformations. Furthermore, even when the vertex-to-vertex map is \emph{discrete conformal}~\cite{springborn2008conformal}, the corresponding piecewise linear map can induce very large angular distortions (see Fig.~\ref{fig:pl_is_bad}).

We propose an alternative triangle-to-triangle map, denoted \emph{blended piecewise \mobius} (BPM), which is based on \mobius transformations, and leads to considerably less artefacts. First, when the vertex-to-vertex map is discrete conformal, BPM yields a low quasi-conformal distortion. Furthermore, BPM is equivariant to global \mobius transformations, and is \mobius transformation reproducing. This allows us to define BPM between surfaces and planar meshes, by defining the map \emph{locally}. Finally, BPM is applicable to \emph{any} vertex-to-vertex map, and leads to smoother texture transfer compared to the alternatives.

\subsection{Related work}
There is a large number of works on computing conformal maps, whether approximated, e.g.,~\cite{vaxman2015conformal,sawhney2017boundary}, under some definition of discrete conformality, e.g.~\cite{springborn2008conformal}, or defined smoothly on the domain e.g.~\cite{weber2009complex,weber2011complex}. 

Our work, however, deals with the \emph{interpolation} of a given \emph{discrete} map, to a smooth map with different properties.
To the best of our knowledge, there are very few such interpolators. Of course, one can use a smooth conformal~\cite{weber2009complex} or quasi-conformal~\cite{weber2011complex} map, and add constraints for the interpolated vertices.
However, such an approach will often lead to over constrained systems, which either do not interpolate the constraints, or create double covers.

In terms of \emph{local} interpolators, it is possible to use a piecewise-linear map; however, it leads to visible artefacts for coarse triangulations. Furthermore, our goal is to design an interpolator that commutes with \mobius transforms, and of course, a linear (or higher order) map will in general not have this property. Finally, it is possible to use a projective interpolation scheme~\cite{springborn2008conformal,born2017quasiconformal, Gillespie2021DCE}. This approach leads to nice results when applied to \rev{discrete conformal maps;}
however it is \emph{discontinuous} on general deformations. 

\rev{We note that some methods~\cite{Crane2011, Chern:2015} approached conformal mappings by designing a \emph{discretized}, rather than \emph{discrete} (cf.~\cite{vaxman2015conformal}) field of rotations and scale factors that were integrated into a map which was conformal up to integrability. Specifically,~\cite{Chern:2015} constructed a representation of this field in volumes that by itself construes an interpolation of \mobius maps. However, these works did not explicitly present a continuous and interpolating blend for triangle meshes as we do.}






\subsection{Contributions}
Our main contributions are:
\begin{itemize}
    \item BPM: A vertex-interpolating, non-linear triangle-to-triangle map, which is smooth across triangles.
    
    \item BPM is equivariant to \mobius transformations, and has low quasi-conformal distortion when the vertex-to-vertex map is discrete conformal.
    
    \item BPM provides a smooth texture pullback, even for very coarse triangulations, and for \emph{any} vertex-to-vertex map.  
\end{itemize}
\section{Background}

We describe our method first as a plane-to-plane map in global planar coordinates, and show how it is easily generalizable to curved surfaces with local intrinsic coordinates in Section~\ref{sec:curved-surfaces}.

\subsection{Discrete and continuous maps}
Consider a triangle mesh $\element{M}=\left\{\element{V},\element{E},\element{T}\right\}$, embedded in the complex plane $\mathbb{C}$ without overlaps. We parameterize the embedding by the vertex coordinates,  $\coord{Z}=\left\{\coord{z}_v \in \mathbb{C} \,|\, v \in \element{V}\right\}$. A map $\dmap{F}:\coord{Z} \rightarrow \coord{W}$, which transforms the vertex positions by $\dmap{F}(\coord{z}_v) = \coord{w}_v$, is denoted \emph{discrete}. We are mainly interested in computing an \emph{interpolation} of a discrete map $\dmap{F}$ into a \emph{continuous} map $f:\cmap{Z}\rightarrow\mathbb{C}$, where $\cmap{Z}$ is the union of all the triangles defined by $\element{T}$ with vertex coordinates in $\coord{Z}$. Such a map is \emph{interpolating} when $\forall v \in \element{V},\ f(z_v)=\dmap{F}(\coord{z}_v)$. We define the \emph{interpolator} as the operator $o:(\cmap{Z},\dmap{F})\rightarrow\mathbb{C}$, such that:
$$
f(z) = o(z, \dmap{F}).
$$
For instance, barycentric interpolation is an interpolator that generates piecewise-linear functions.

\subsection{Holomorphic maps} A differentiable map $f\!:\! \xR^2\! \to\! \xR^2$, $f=(u(x,y),v(x,y))$ with  a Jacobian of the form $\nabla f =\big( \begin{smallmatrix}a & b \\ -b & a\end{smallmatrix}\big)$, is \emph{holomorphic}, when considered as a function on the complex plane, $f\!:\! \xC\! \to\!\xC$, where $f(x+iy) = u(x,y) + iv(x,y)$. Alternatively, this can be written as $\frac{\partial f}{\partial \conj{z}}=0$, indicating that a complex function that is independent of $\conj{z}$ is holomorphic. Holomorphic maps preserve the angle between any two intersecting curves, and are therefore detail preserving and useful for texture mapping. A simple example of a holomorphic map $f\!:\! \xC \! \to\! \xC$ is the complex affine map $f(z) = az+b$, for some $a,b\tin\xC$, which is a global similarity transformation (i.e., scale, rotation and translation). Such a map is uniquely defined by the transformation of two points. 

Perhaps the quintessential holomorphic map is the \emph{\mobius transformation} (defined on the extended complex plane $\xCe = \xC \cup {\infty}$), which has the form $m(z)=\mob{a}{b}{c}{d}$, for some $a,b,c,d\tin\xC$ such that $ad-bc \neq 0$. The parameters $a,b,c,d$ are unique up to a multiplicative factor $\alpha\tin\xC$. We therefore additionally assume the normalization $ad-bc=1$, which leads to uniqueness of the parameters up to sign. By working with complex homogeneous coordinates, a \mobius transformation $m(z)$ can also be represented as a matrix $M\!=\!\big( \begin{smallmatrix}a & b \\ c & d\end{smallmatrix}\big)\tin\xC^{2\times2}$ with determinant $1$. Then, we have $M [z ; 1] = [az+b ; cz+d] \equiv [m(z) ; 1]$. The matrix representation of the composition of two \mobius maps $m_1( m_2(z) )$ is given by the multiplication of their matrix representations, i.e., by $M_1 M_2$. Similarly, the matrix representation of $m^{-1}$ is $M^{-1}$. \mobius maps include similarities and inversions in spheres, and are defined uniquely by the transformation of \emph{three points}. Since both $M$ and $-M$ represent the same transformation $m$, we use $\equiv$ to denote matrix equality up to sign, i.e. $M\equiv -M$. The choice of the sign is only required when taking a unique root or logarithm of a \mobius matrix, as elaborated in Sec.~\ref{subsec:mobius-interpolator}. 

Barycentric blends of complex affine maps have been used successfully for generating interpolators for \emph{polygonal domains}~\cite{weber2011complex}, by blending the complex affine maps defined by the deformation of the polygon \emph{edges}. 
We generalize this idea, and propose to use \emph{blends of \mobius maps} for generating an interpolator for a discrete map between two planar \emph{triangle meshes}, by blending the \mobius maps defined by the deformation of the \emph{triangles}. 




\subsection{Piecewise-Compatible \mobius Maps}
\label{sec:Piecewise-Compatible}

We parameterize any discrete map $\dmap{F}\!:\!\coord{Z}\!\to\!\coord{W}$ with a
set of \mobius transformations $\left\{\dmap{m}_t \, | \, t=(i,j,k) \tin \mT\right\}$ defined uniquely per triangle by the transformation of the vertices: $m_t(z_i)=w_i,m_t(z_j)=w_j,m_t(z_k)=w_k$. We denote by $\{M_t \tin \xC^{2\times 2} \,|\, t\tin\mT\}$ the corresponding matrices, with components $a_t, b_t, c_t, d_t \tin \xC$.


\paragraph*{Compatibility condition.} A set of transformations $\left\{\dmap{M}_t\right\}$ is \emph{compatible} with a map $F\!:\!Z\to W$ if the transformations of neighboring triangles agree on the map of their \emph{common vertices}. Specifically, given two adjacent triangles $t_1=(i,j,k),t_2=(j,i,l) \tin \mT$ with a shared edge $e=(i,j)$, we have that $w_i = \dmap{M}_{t_1}(\coord{z}_i) = \dmap{M}_{t_2}(\coord{z}_i)$ and similarly for $\coord{z}_j$. 

Given a triangle mesh $\mM$, a set of \mobius transformations $\{M_t\}$ that fulfills the compatibility condition defines a \emph{Piecewise-Compatible \mobius (\textbf{PCM}) Map} ~\cite{vaxman2015conformal}. It is advantageous to consider general deformations as PCMs (as opposed to, e.g., piecewise-affine maps) due to their natural connection to conformal and discrete conformal deformations. For example, PCM maps are closed under global (single) \mobius transformations. Namely, given a matrix representation $M_g$ of a global \mobius transformation $m_g$, we have that the set of transformations $\{M_t M_g\}$ and $\{M_g M_t\}$ are also PCM maps. In addition, discrete conformality (CETM)~\cite{springborn2008conformal} has an elegant description in the PCM representation in terms of the \emph{corner variables} $\{X_{t,i}\tin\xC \,|\, t\tin\mT, i\tin t, v_i \tin\mV\}$, where $X_{t,i}=(c_t z_i + d_t)^{-1}$. Specifically, a PCM map is a discrete conformal equivalence if and only if $|X_{t,i}|$ does not depend on $t$. Then, $|X_{\cdot,i}|=e^{u_i/2}$, where $u\!:\!\mV\to\xR$ is the conformal factor.


Unfortunately, unlike the piecewise-affine interpolation, the trivial interpolation of a discrete PCM map, where the \mobius transformation $\dmap{M}_t$ is applied to every point $z \in t$, is not continuous between triangles. A simple way to see this is that a \mobius map is uniquely determined by $3$ points. Therefore, the transformation of \emph{all the points on the edge} shared by two triangles is compatible by both triangles if and only if they are transformed by a single \mobius transformation, which means that the entire mesh is. Our challenge is then to find an \emph{interpolator} of PCM maps.

\section{Blended Piecewise \mobius Maps}
\subsection{Blended Maps Desiderata}
\label{sec:desiderata}
 Given an input discrete map $\dmap{F}\!:\!Z\!\to\! W$, denote by $\dmap{M}(\dmap{F})=\{M_t\,|\,t\in\mT\}$ the PCM map (i.e., the \mobius matrices) induced by $\dmap{F}$.
We define a \emph{map} interpolator $o(\cmap{Z},F)$ using a continuous \emph{\mobius matrix} interpolator $O\!:\!(\cmap{Z},M(F)) \!\rightarrow\! \mathbb{C}^{2 \times 2}$, namely a \mobius transformation $O(z,M(F))$ with spatially varying blended coefficients. We then define $O$ and $o$ such that:
 \begin{equation}
 [o(z,F);1] \equiv O(z,M(F))[z;1].     
 \end{equation}

 Our requirements from the PCM interpolator $O(z,\dmap{M})$ of $\dmap{M}$ are:
\begin{enumerate}
\item \textbf{Locality.} $O(z,M)$ should depend only on the local neighborhood of $z$.
\item \textbf{Identity reproduction.} $O(z,\{M_t \equiv Id\})\equiv Id$.
\item \textbf{Continuity.} The resulting map $o(z,F)$ should be at least $C^0$-continuous between neighboring triangles.
\item \textbf{\mobius equivariance.} The interpolator should commute with \mobius transformations. That is, for any global \mobius transformation $M_g$ we have:
\begin{align}
O(z, \{\dmap{M}_g \dmap{M_t}\}) &\equiv  \dmap{M}_g O(z, \dmap{M}). \nonumber \\
O(z, \{ \dmap{M_t} \dmap{M}_g \}) &\equiv   O(z, \dmap{M})\dmap{M}_g.
\label{eq:mobius-commutation}
\end{align}
Namely, interpolating the discrete map and performing a global \mobius transformation can be done in any order for the same result. 
\item \textbf{\mobius reproduction.} If all vertices are transformed by \emph{the same} \mobius transformation $\dmap{M}_g$ then the interpolator $O$ reproduces that \mobius transformation, i.e., $O(z,\dmap{M}_g) \equiv \dmap{M}_g$. This is a corollary of Properties (2) and (4). 

\end{enumerate}

\rev{We note that \mobius equivariance is essential for the consistency of interpolating CETM maps; the set of CETM maps are closed under \mobius transformations;  specifically, any global \mobius transformation induces a CETM map. Properties (4) and (5) then guarantee that this property carries over to our interpolator.}

\rev{We prove in Sec.~\ref{subsubsec:properties}} that our requirements are met by the interpolator that we define in Sec.~\ref{subsec:mobius-interpolator}. We further list objectives for the interpolator that we empirically witnessed in all our examples:

\begin{enumerate}
\item \textbf{CETM interpolation.} If the interpolator is applied to a CETM map $\dmap{M}$, then the result should be a close approximation to a continuous conformal map.
\item \textbf{QC Errors are bounded.} The quasiconformal error of the interpolated $M(z)$ for any $z \in t \in \element{T}$ is bounded above by the (discrete) quasiconformal error of $t$ in $\dmap{M}$.
\end{enumerate}
We list the above as objectives since we do not have explicit proofs that they are always true; nevertheless we provide ample empirical evidence in Sec.~\ref{sec:results}. 

\subsection{\mobius Interpolator}
\label{subsec:mobius-interpolator}
\subsubsection{The \mobius ratio} 
\begin{figure}
\centering
\includegraphics[width=0.2\textwidth]{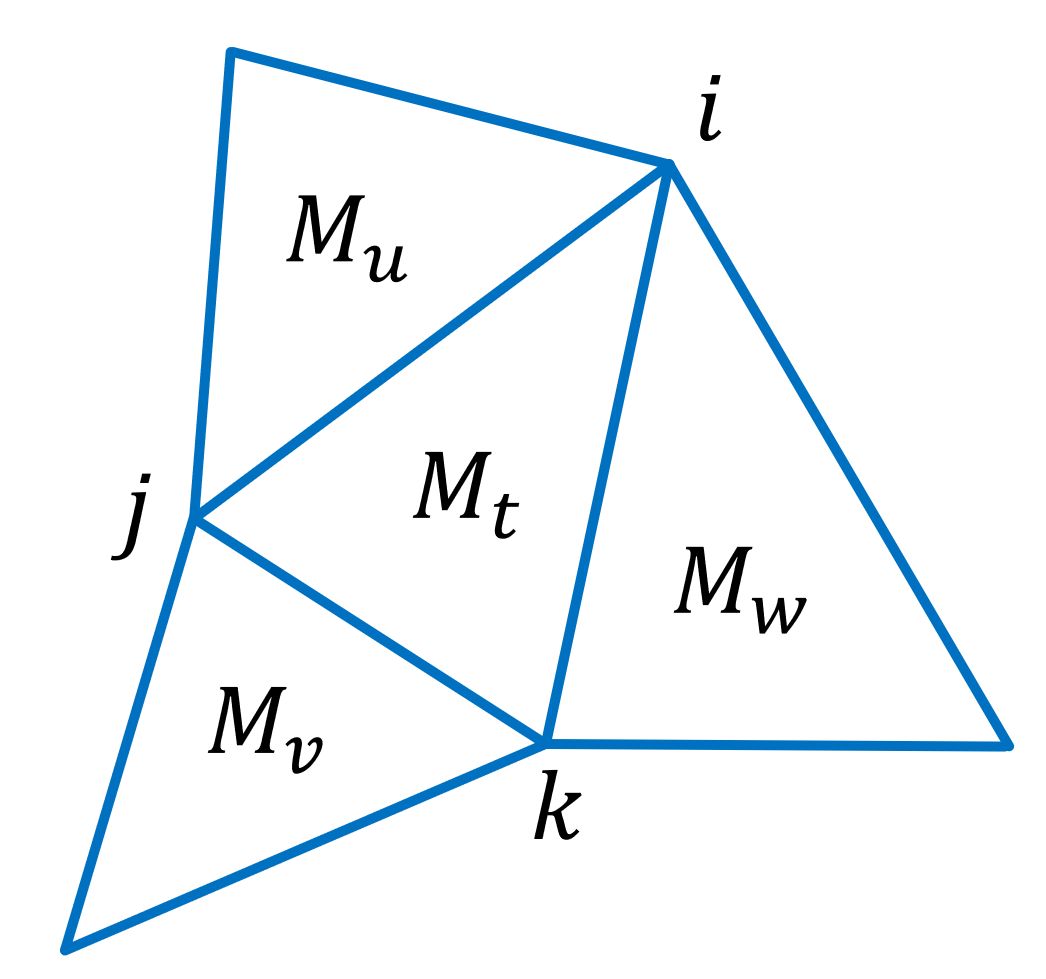}
\caption{Our notation.}
\label{fig:notation}
\end{figure}

Let $M_t, M_u \in \xC^{2\times2}$ be two normalized \mobius matrices representing transformations on two faces adjacent at edge $e_{ij}$ (see Fig.~\ref{fig:notation}). The \emph{\mobius ratio} $\md_{tu}$ is given by:
\begin{equation}
    \md_{tu} = M_t M_u^{-1}.
\label{eq:delta}
\end{equation}
Intuitively, the \mobius ratio describes the difference between applying $M_u$ and applying $M_t$, in the sense that $M_t = \md_{tu} M_u$. 
It is easy to check that $\md_{tu}^{-1} \equiv \md_{ut}$, and $\md_{tu}\equiv Id$ if and only if $M_t \equiv M_u$. 
Furthermore, due to the PCM compatibility between $M_t$ and $M_u$, we have that $F(z_i)$ and $F(z_j)$ are \emph{fixed points} of the transformation $\md_{tu}$.

We additionally define the \emph{log \mobius ratio}, given by:
\begin{equation}
\lmd_{tu} = \log\left(\text{Sign}(\text{Tr}(\Re (\md_{tu}))) \cdot \md_{tu}\right),
\label{eq:log-mobius-ratio}
\end{equation}
\rev{where $\Re()$ is the real part of a complex number, $\text{Tr}()$ is the trace operator, and $Sign()$ is the sign of a real number (outputting $\pm 1$).}

Thus, $\lmd_{tu}$ is the log of either $\md_{tu}$ or $-\md_{tu}$, whichever is closer to the identity \rev{in the Frobenius norm} (see Appendix B). 
The square root of the \mobius ratio is correspondingly given by: $\sqrt{\md_{tu}} = \exp(\tfrac{1}{2}\lmd_{tu})$.
 \paragraph*{Boundary edges.} 
 If $e_{ij}$ is a boundary edge, then we set its ratio to Id.
 That encodes the choice that the transformation ``beyond'' the edge is the same \mobius transformation of $t$, which naturally adheres to our requirements.

\subsubsection{Ratio interpolator} \label{subsec:Ratio-interpolator}
 Consider a face $t=ijk \tin \element{T}$ and neighboring triangles $u,v,w\tin\mT$ adjacent to the edges $e_{ij},e_{jk},e_{ki}\tin\mE$, respectively (Fig.~\ref{fig:notation}). 
Each face has a corresponding \mobius matrix $M_t, M_u, M_v, M_w$, and each edge has a corresponding log \mobius ratio of its neighboring triangles: $\lmd_{ut}$, $\lmd_{vt}$ and $\lmd_{wt}$. We define the \emph{log ratio interpolator} as:
\begin{equation}
\lmd_{t}(z,M) = \frac{\bc{ij}(z)\lmd_{ut}+\bc{jk}(z)\lmd_{vt}+\bc{ki}(z)\lmd_{wt}}{\bc{ij}(z)+\bc{jk}(z)+\bc{ki}(z)},
\label{eq:ratio-interpolator}
\end{equation}
for some \emph{edge barycentric coordinates} $0 \leq \bc{e}(z) \leq 1$, with $e \tin \mE_t=\{e_{ij}, e_{jk}, e_{ki}\}$. We require that for $e\!,\!\tilde{e}\tin\mE_t$, and a non-vertex point $z\tin{\tilde{e}},z\!\notin\!\{z_i,z_j,z_k\}$ we have that $\bc{e}(z)/\sum_{\hat{e}\in\mE_t} B_{\hat{e}}(z)=1$ if $e\!=\!\tilde{e}$ and $0$ otherwise. In addition, we require that the sum of the coordinates does not vanish. Specifically, we take $\bc{e}(z) = {d(z,e)}^{-1}$, where $d(z,e)$ is the distance of $z$ to the line the edge $e$ lies on. See Appendix A for the implementation details.

Finally, our \emph{\mobius interpolator} is given by:
\begin{equation}
O(z\in t, M) = \exp\left(\frac{1}{2}\lmd_{t}(z,M)\right)M_t = \sqrt{\md_{t}(z,M)}\,M_t.
\label{eq:mobius-interpolator}
\end{equation}

\paragraph*{Discussion.}
Our interpolator is similar in spirit to the rotation interpolant of Alexa~\cite{alexa2002linear}, and is based on the general approach of interpolation in Lie groups~\cite{marthinsen1999interpolation}. 
By linearly interpolating the \emph{log} \mobius ratio, we guarantee that the blended \rev{matrix $O(z,M)$} is normalized (i.e., has determinant $1$) if the input \rev{matrices $M$} are normalized. That is because the zero-trace property is invariant under a linear blend.

\subsubsection{Properties}
\label{subsubsec:properties}
Our interpolator is local (Req. (1)) since it is defined using a triangle and its $3$ neighbors, and it is easy to check that it reproduces the identity (Req. (2)). 

\paragraph*{Continuity on edges.} 
Without loss of generality, when $z \in e_{ij}, z\neq z_i, z_j$, we have that $\lmd_t(z) = \lmd_{ut}$ and $\lmd_u(z)=\lmd_{tu}$, and thus our interpolation reduces to:
\begin{equation}
O(z, \dmap{M}) = \sqrt{\md_{ut}} \cdot M_t \equiv  \sqrt{\md_{tu}} \cdot M_u , \quad \forall z\in e_{ij}, z\neq z_i,z_j
\label{eq:edge-interpolator}
\end{equation}
Hence, the \mobius interpolator on the edge $e_{ij}$ only depends on the two faces $t,u$ adjacent to the edge, and it is symmetric in $t,u$ (up to sign) leading to the same map $O(z,M)$. 
Note that Eq \eqref{eq:edge-interpolator}
is similar to SLERP interpolation for quaternions~\cite{shoemake1985animating}. 

\paragraph*{Continuity on vertices.}
Note that the barycentric coordinates are not continuous on a vertex (e.g. $z_i$), hence the ratio interpolant is also not continuous at the vertex. However, we have that $F(z_i)$ is a \emph{fixed point} of the \mobius ratios, and thus we interpolate the original PCM map at $z_i$. This leads to continuity on vertices across different triangles, as needed by Req. (3). 




\paragraph*{M\"{o}bius equivariance.} We first note that the ratios $\md$ are invariant to right composition $M_{t|u|v|w}M_g$ with a global \mobius transformation $M_g$; thus, the interpolant $O$ is trivially equivariant to right composition. For left composition $M_gM_{t|u|v|w}$ (first PCM then global), we have a conjugated ratio $\md_{tu}=M_g(M_tM_u^{-1})M_g^{-1}$. Since trace is invariant to cojugation, and since conjugation commutes with matrix logarithm and exponent, the entire interpolant becomes:
\begin{equation}
O(z \in t, \{M_g M_t\})=M_g \md(z,\{M_t\}) M_g^{-1}\cdot  M_g M_t=M_g O(z\in t, M).    
\end{equation}
Thus, we also fulfill Req. (4), and with (2) we fulfill Req. (5). 

\paragraph*{Local injectivity.} \mobius transformations are locally injective in a region that does not contain poles. Specifically, if a single \mobius transformation $m_t$ of a triangle $t$ does not flip or degenerate the triangle edges, we have that $m_t$ has a positive Jacobian anywhere inside. Nevertheless, for the \emph{blended} \mobius transformation we do not have such a guarantee. In practice, our maps are well behaved for the blending weights that we have chosen, \rev{however extreme cases may exist (see Figure~\ref{fig:local-injectivity})}.

\section{Curved surfaces}
\label{sec:curved-surfaces}
Our method is also applicable for mapping from curved surfaces to the plane. 
The discrete mapping is computed \emph{locally} for each triangle, by flattening it and its neighboring three triangles isometrically to the plane to generate the source triangles $Z$.
The continuous mapping is then computed by blending inside the triangle, using the same scheme as in the two-dimensional case, and pulling the resulting map back to the surface.

More formally, consider a triangle mesh $\element{M}=\left\{\element{V},\element{E},\element{T}\right\}$, embedded in $\mathbb{R}^3$. Let  $\coord{X}=\big\{\coord{x}_v \in \mathbb{R}^3 \,|\, v \in \element{V}\big\}$  be its vertex coordinates. The discrete map $\dmap{F}:\coord{X} \rightarrow \coord{W}$ transforms the vertex positions by $\dmap{F}(\coord{x}_v) = \coord{w}_v \in \mathbb{C}$.
We are interested in computing a continuous interpolating map $f:\cmap{X}\rightarrow\mathbb{C}$, where $\cmap{X}$ is the union of all the triangles defined by $\element{T}$ with vertex coordinates in $\mathbb{R}^3$ and $\forall v \in \element{V},\ f(x_v)=\dmap{F}(\coord{x}_v)$. 

We define for each $t \tin \mT$, a \emph{local} discrete map $\tilde{\dmap{F}_t}:\tilde{\coord{Z}}_t \rightarrow \coord{W}$ where $\tilde{\coord{Z}}_t$ is an isometric embedding in 2D of the face $t$ and its neighboring faces $u,v,w$. The corresponding \mobius matrices $M(\tilde{F_t}) = \{ M_t, M_u, M_v, M_w\}$ are defined as before, as is the matrix interpolator $O(z, M(\tilde{F}_t))$, and correspondingly the interpolator $o(z, \tilde{F}_t)$.
\rev{Let $\tilde{\coord{z}}_t \tin \mathbb{C}$} be the planar point that corresponds to some point $x \in t$ \rev{on the mesh} under the local isometric embedding.
The interpolator is defined $\forall t\in\element{T}$ as follows:
\rev{
\begin{equation}
f_t(x) = f_t(\tilde{\coord{z}}_t) = o(\tilde{\coord{z}}_t, \dmap{\tilde{F_t}}).
\end{equation}
}


\subsection{Continuity} 
We need to show that this definition is well-posed, since it is defined for each triangle separately.
We get this since (1) Our interpolator is \mobius equivariant, (2) there exists a \mobius map between isometric embeddings, and (3) the map of points on the edge depends only on the \mobius matrices of its neighboring triangles. 


\begin{figure}
    \centering
   \includegraphics[width=0.7\linewidth]{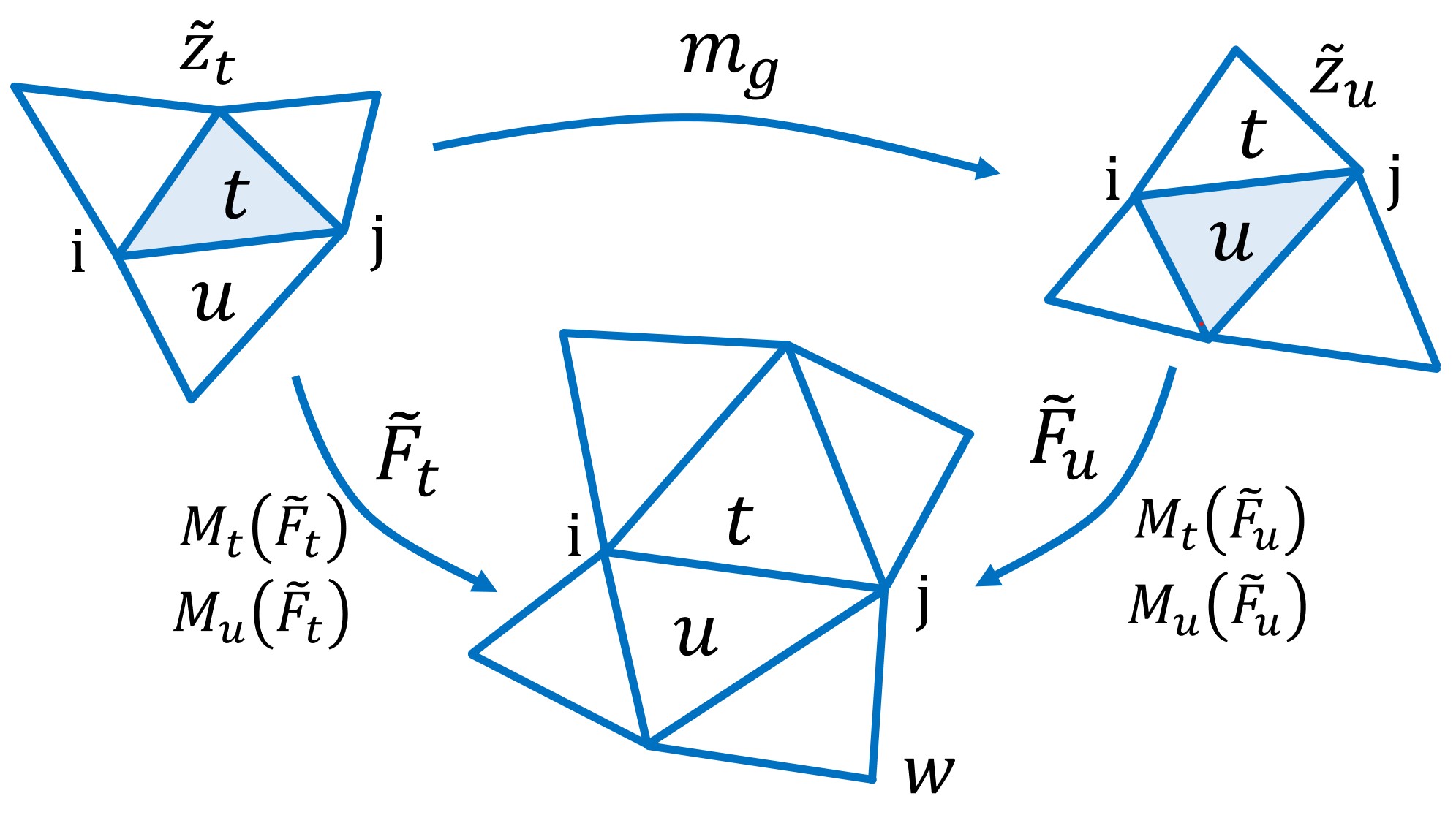}
    \caption{Notation for 3D Framework.}
    \label{fig:illustration_3D}
\end{figure}

Formally, Let $t,u \tin \mT$, be two triangles that share an edge $e_{ij}$, and let $\tilde{\dmap{Z}}_t,\tilde{\dmap{Z}}_u$ be the corresponding (independent) isometric embeddings of each triangle and its neighboring faces. See Fig.~\ref{fig:illustration_3D} for our notation. \rev{Since the two embeddings map the triangles $t,u$ isometrically to the plane, there exists a \mobius transformation $m_g$ such that $\forall x \tin {t \cup u}$, its corresponding planar points $\tilde{z}_t \in \tilde{Z}_t$ and $\tilde{z}_u \in \tilde{Z}_u$ satisfy $\tilde{z}_u = m_g(\tilde{z}_t)$. }
We denote by $M_t(\tilde{F}_t), M_t(\tilde{F}_u)$ the \mobius matrices corresponding to $t$ induced by $\tilde{F}_t, \tilde{F}_u$, respectively, and similarly for $M_u(\tilde{F}_t), M_u(\tilde{F}_u)$. By construction, we have that: 
\begin{equation}
    M_t(\tilde{F}_t) \equiv M_t(\tilde{F}_u) M_g, \quad\quad M_u(\tilde{F}_t) \equiv M_u(\tilde{F}_u) M_g,
\end{equation}
where $M_g$ is the \mobius matrix that corresponds to $m_g$. 

Let $x\tin e_{ij}$ be a point on the mutual edge of $t$ and $u$, with the corresponding planar points \rev{$\tilde{z}_t, \tilde{z}_u$}. 
The interpolator of a point on the edge depends \emph{only} on the \mobius matrices of its neighboring triangles, and is given by Equation~\eqref{eq:edge-interpolator}. We have:
\begin{equation}
    \md_{tu}(\tilde{F}_t) = M_t(\tilde{F}_t) M_u^{-1} (\tilde{F}_t) = M_t(\tilde{F}_u) M_g M_g^{-1} M_u^{-1}(\tilde{F}_u) = \md_{tu}(\tilde{F}_u).
\end{equation}

Thus, the matrix interpolator is given by
\begin{equation}
\begin{aligned}
    O(\tilde{z}_t, M(\tilde{F}_t)) &= \sqrt{\md_{ut}(\tilde{F}_t)} M_t(\tilde{F}_t) =\\
                                   &= \sqrt{\md_{ut}(\tilde{F}_u)} M_t(\tilde{F}_u) M_g = O(\tilde{z}_u, M(\tilde{F}_u)) M_g.    
\end{aligned}
\end{equation}
Finally, we have:
\begin{equation}
\begin{aligned}
    [o(\tilde{z}_t, \tilde{F}_t);1] &= O(\tilde{z}_t, M(\tilde{F}_t))[\tilde{z}_t;1] =\\
                                   &= O(\tilde{z}_u, M(\tilde{F}_u)) M_g M_g^{-1} [\tilde{z}_u;1] = [o(\tilde{z}_u, \tilde{F}_u);1].    
\end{aligned}
\end{equation}

Hence, we have that the map interpolation is consistent, as required. Note that this consistency generalizes to \emph{any} locally defined interpolator, as long as it is equivariant to maps between the local flattened patches. 
We present results in Fig.~\ref{fig:teaser} and in Sec.~\ref{sec:results}.
\rev{Note that our map is at least $C^0$ continuous, but not $C^1$ in general}.

\rev{We provide the pseudo code for our algorithm in Appendix C.}
\begin{figure}[b]
    \centering
    \includegraphics[width=\linewidth]{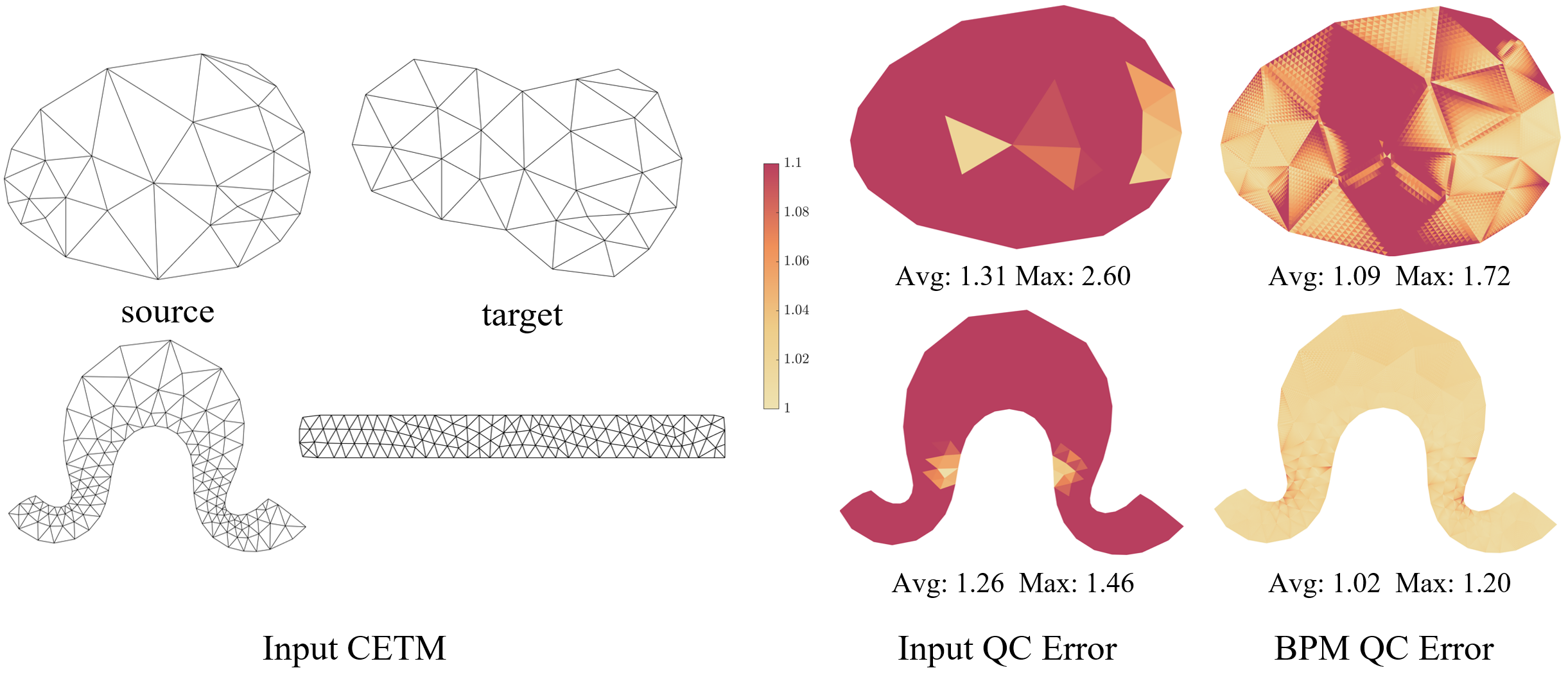}  
    \caption{CETM as input. (left) The input CETM deformation. (right) The QC errors of the input discrete deformation and the BPM mapping. Note that the error of BPM is considerably lower than the input errors.}
    \label{fig:CETM_input}
\end{figure}
\section{Experimental Results}
\label{sec:results}
We use a variety of examples to demonstrate the effectiveness of our interpolators. 
For each example, we show the source and target meshes, and visualize the map by (1) pulling back a texture from  the target mesh to the source mesh, as well as (2) pushing forward a texture from the source mesh to the target mesh. Note that on the target mesh, the edges are \emph{curved}.
While our interpolator is smooth and in closed-form, computing the resulting Quasi-conformal (QC) distortion introduces a complicated expression which varies non-linearly within the triangle. To facilitate its visualization, we simply approximate the resulting QC error by refining the source mesh using $4$ levels of subdivision, applying the computed (continuous, non-linear) interpolator to the refined vertices, and computing the QC distortion of the linear map between the subdivided triangles. For a single subdivided triangle, the QC distortion is given by the ratio of the singular values of the linear map~\cite{sander2001texture}.

For the input discrete deformations we use different deformations/parameterization techniques. We use Conformal Equivalence of Triangle Meshes (CETM) 
~\cite{springborn2008conformal} and Boundary-First Flattening (BFF)~\cite{sawhney2017boundary} 
for generating discrete conformal input maps. For pure planar deformations, we use As-\mobius-as-possible (AMAP)~\cite{vaxman2015conformal} for discrete maps with small QC and CETM distortion. We use Cauchy coordinates (CC)~\cite{weber2009complex} to generate discrete deformations sampled from continuous conformal maps.
We additionally use As-Killing-As-Possible shape deformation (AKVF)
~\cite{solomon2011killing} to generate inputs that are far from conformal. 
For additional mappings of surfaces to the plane we use models from the recent parameterization dataset~\cite{shay2022dataset} in Figs.~\ref{fig:comparison_3d_cetm},~\ref{fig:Artist_3D}. The parameterization method used is mentioned in each example.

For comparison, we consider piecewise linear (PL) interpolation, and circumcircle preserving projective interpolation (PROJ)~\cite{springborn2008conformal,born2017quasiconformal}. 

\subsection{Properties}
We first validate the two objectives mentioned in Sec.~\ref{sec:desiderata}.
\paragraph*{CETM as input.}
When the discrete input map is a conformal equivalence, i.e., fulfills the CETM conditions, our interpolator leads to a low QC distortion, even when the QC distortion of the input map is quite large. We demonstrate this for two input deformations in Fig.~\ref{fig:CETM_input}. 

\paragraph*{Bounded QC Errors.}
In all cases the QC error of our map is lower than the QC error of the input map. When the input deformation is close to conformal (Figs.~\ref{fig:comparison_amap},~\ref{fig:comparison_cetm},~\ref{fig:comparison_bff}), our method gives the best results. However, even for deformations far from conformal, (Fig.~\ref{fig:comparison_kvf}), our mapping is smooth with small QC errors.

\subsection{Robustness}
\begin{figure}
    \centering
    \includegraphics[width=1.0\linewidth]{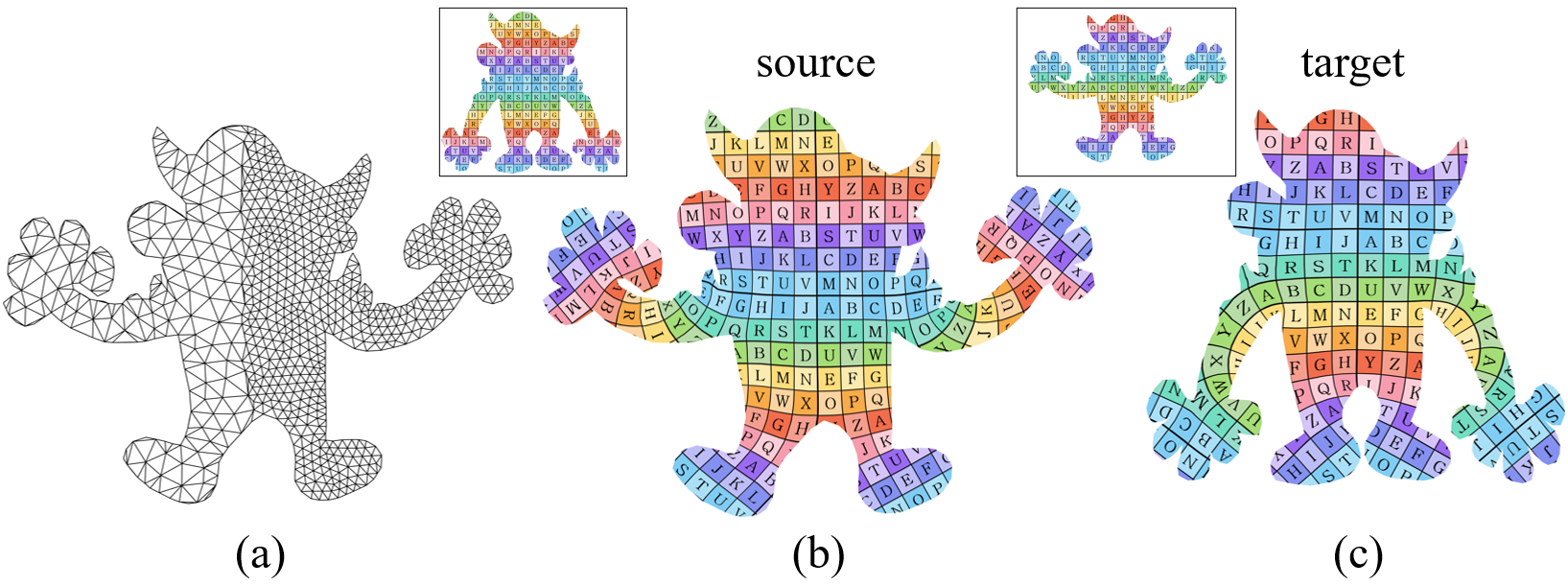}
    \caption{Non-uniform triangulations. (a) The input triangulation, (b) the pullback and (c) push-forward of the texture shown in a black frame with our mapping.} 
    \label{fig:non-uniform}
\end{figure}

We demonstrate the robustness of our approach to different meshes.
\paragraph*{Non-uniform triangulations.} We use a mesh whose left and right halves are meshed differently. We deform it using AKVF, and show the interpolation results in Fig.~\ref{fig:non-uniform}. Note that the texture deformed using our map looks similar on the left and right side of the mesh, thus our method is not sensitive to meshing.

\begin{figure}[b]
    \centering
  \includegraphics[width=1.0\linewidth]{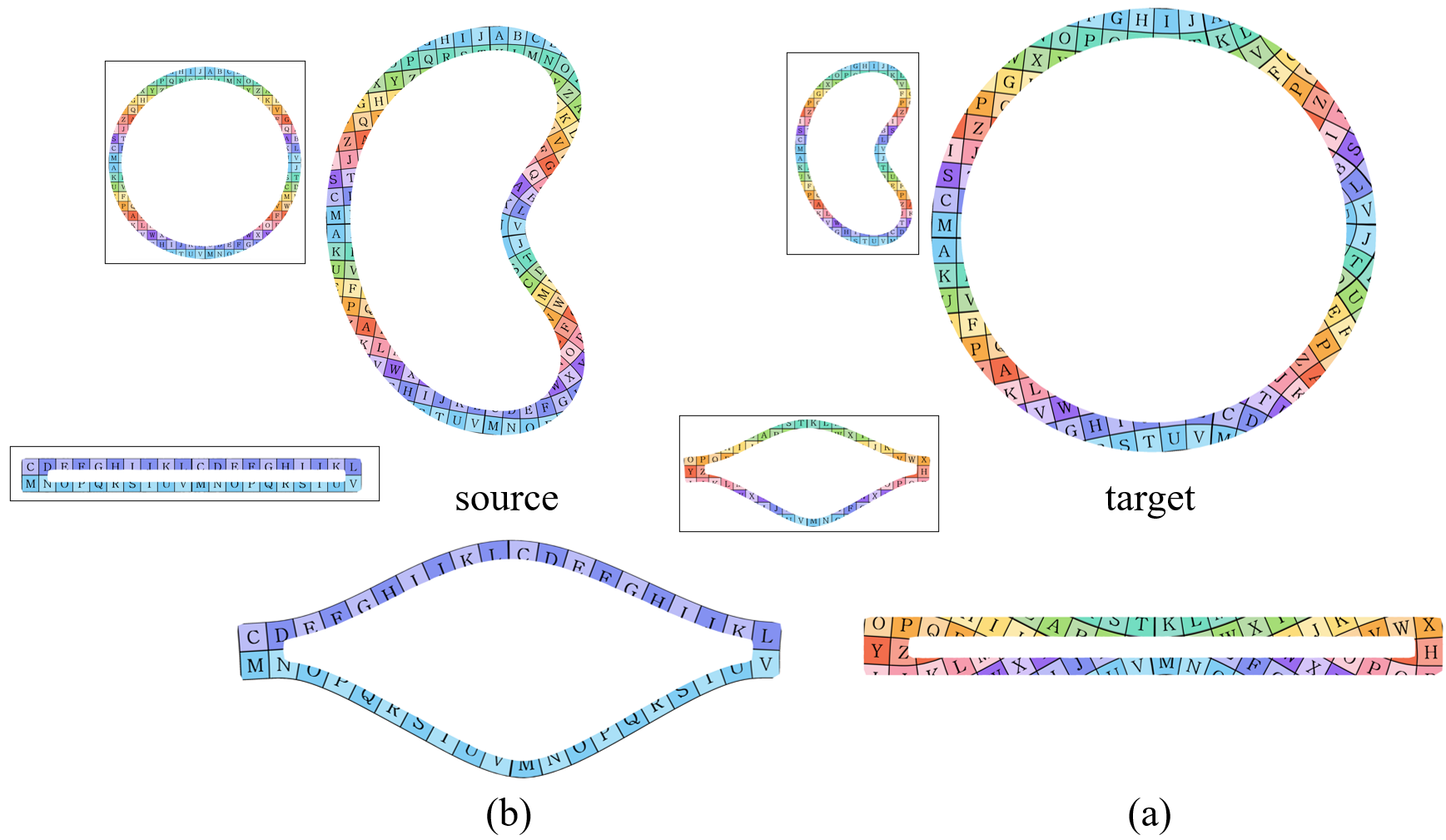}
    \caption{Non simply connected meshes. (a) The pull-back, and (b) the push-forward of the texture (shown in a black frame) using our mapping for two non simply connected meshes. }
    \label{fig:non-simply-connected}
\end{figure}

\paragraph*{Non simply connected.} Our method is applicable to meshes of any topology. We demonstrate it on a few non-simply connected meshes in Fig.~\ref{fig:non-simply-connected}.

\begin{figure}[t]
    \centering
    \includegraphics[width=1.0\linewidth]{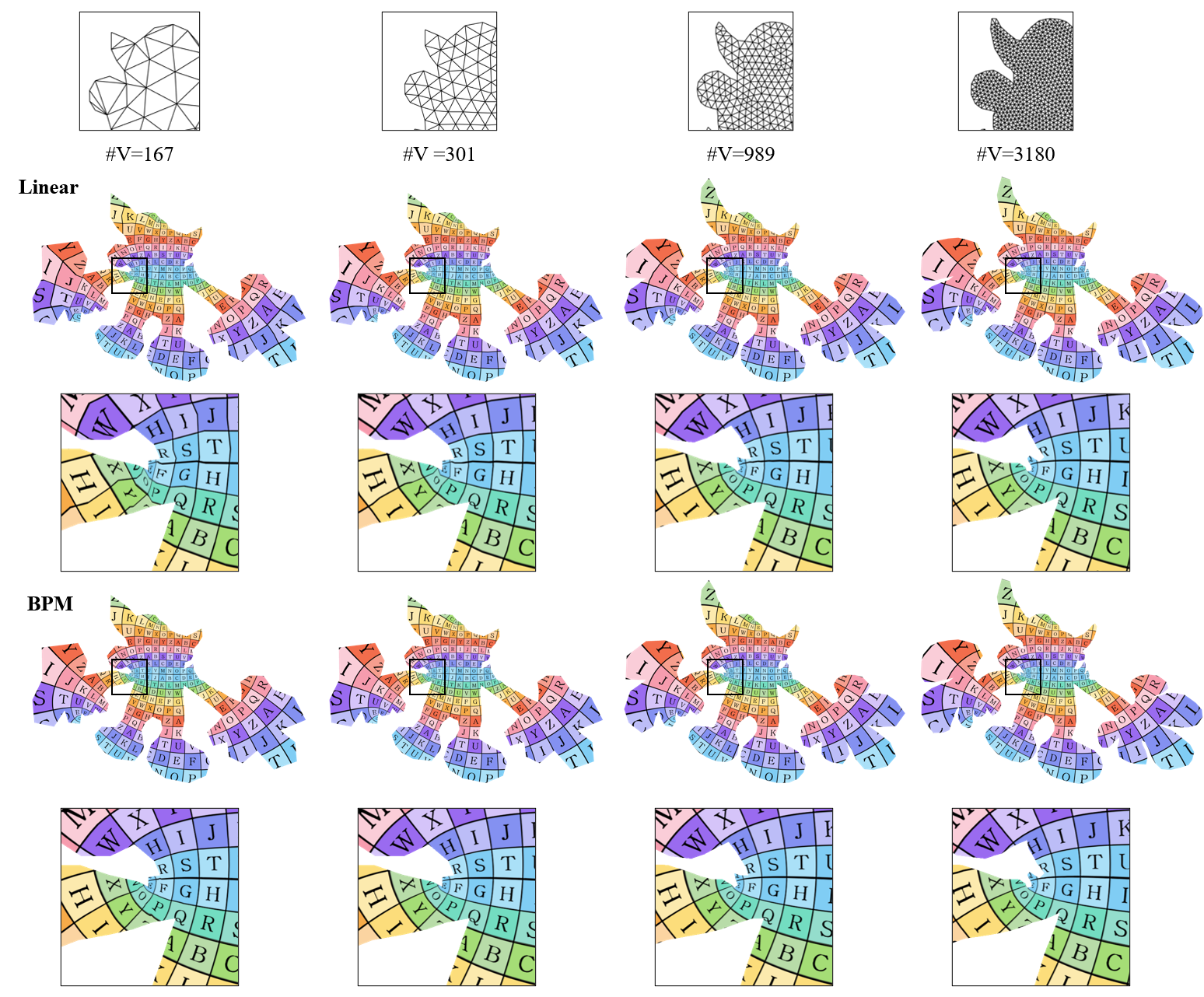}
    \caption{Multiple resolutions. Pull-back of our mapping. from left to right: increased mesh density. Note that our mapping of the \emph{coarse} triangulation (bottom left) is comparable to the linear map on the much denser triangulation (top right).}
    \label{fig:multiple-res}
\end{figure}

\paragraph*{Different resolutions.} 
We remesh a model to $4$ different resolutions, and apply the same deformation by sampling the continuous Cauchy Coordinates, using the same source and target cages. We show the result in Fig.~\ref{fig:multiple-res}, and compare with piecewise-linear interpolation. Note that, unlike the PL map, our results are virtually indistinguishable across resolutions, despite the very different mesh resolutions. 

\begin{figure}[b]
    \centering
    \includegraphics[width=\linewidth]{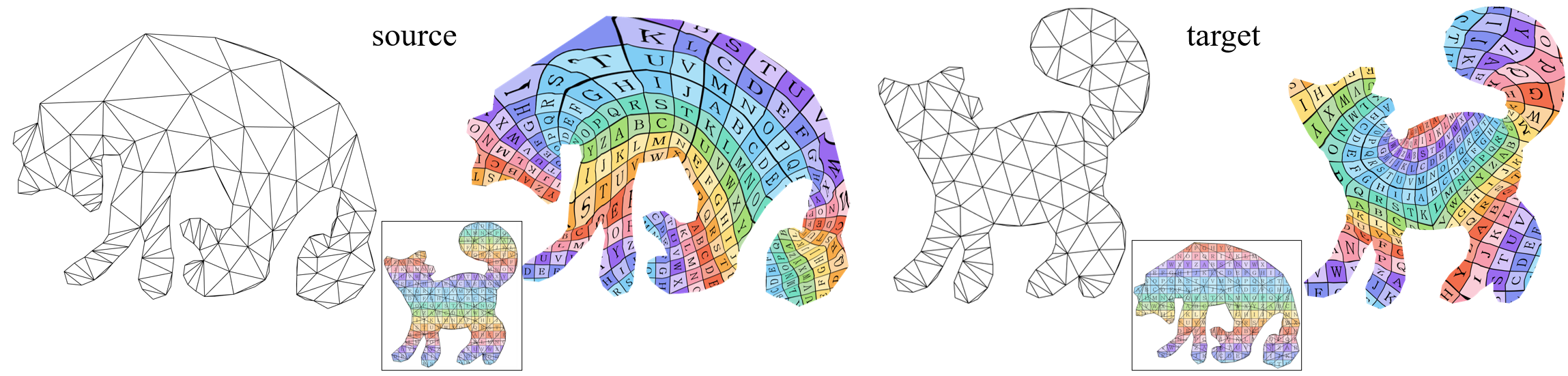}  
    \caption{Even when the input map is far from conformal (here computed using AKVF), our interpolator leads to a smooth map.}
    \label{fig:stress-test}
\end{figure}

\paragraph*{Large deformations.} We assume that the discrete map is slowly varying between triangles, therefore $\md_{tu}$ is close to $Id$ or $-Id$, and the chosen logarithm branch will be the same for the $3$ edges of the triangle. 
However, even if this is not the case, our interpolator is smooth, but may be more oscillatory. 
In this experiment, we demonstrate that our map is resilient to large changes in the deformation of neighboring triangles. In Fig.~\ref{fig:stress-test} we show a discrete map with very large deformations, where our map is still smooth. 

\begin{figure}[t]
     \centering
     \includegraphics[width=\linewidth]{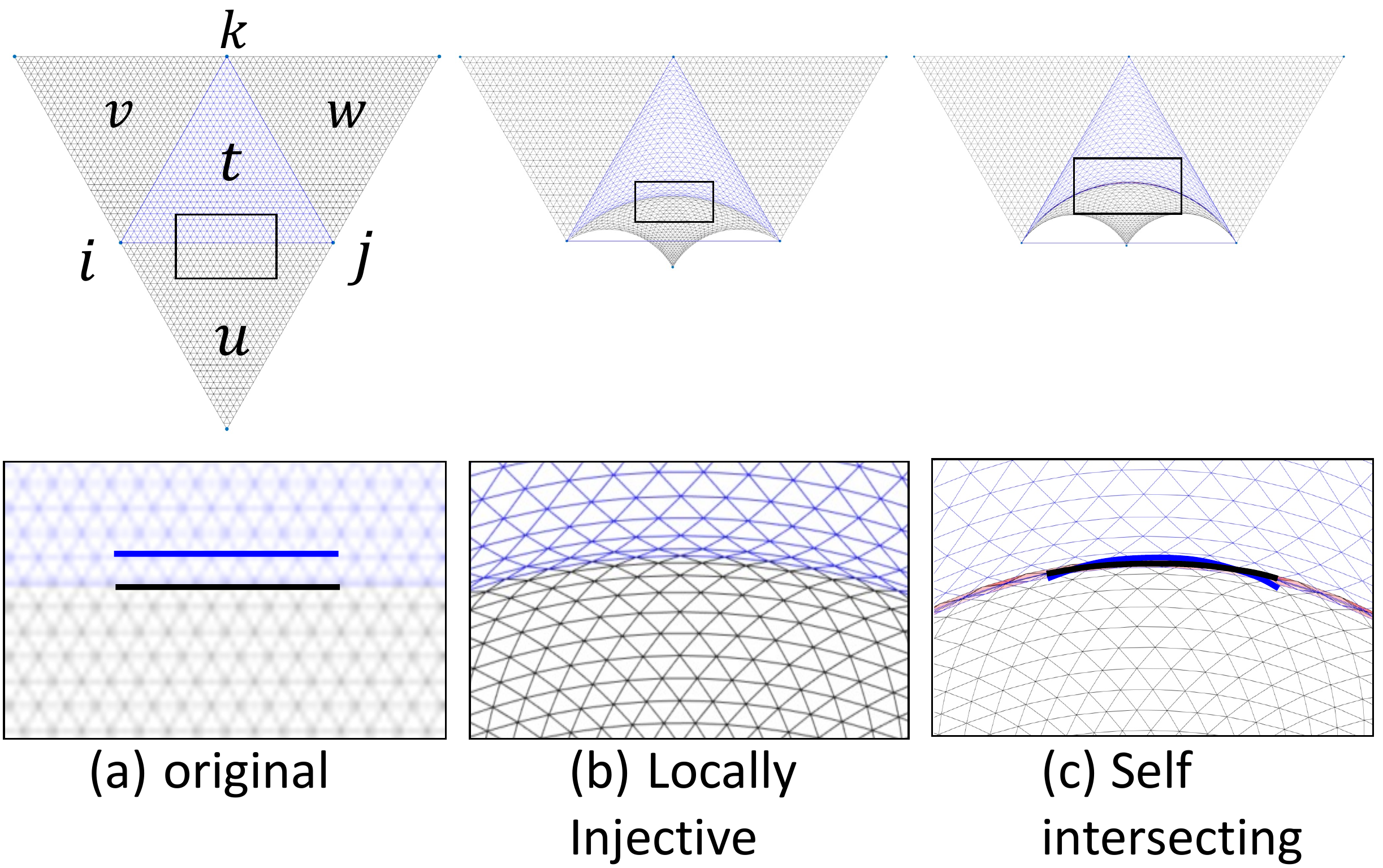}
     \caption{Example of a non-locally-injective transformation. (a): original triangles with part of $e_{ij}$ in black and a parallel line inside $t$ in blue. (b): an extreme deformation with matrix $M$ of the bottom triangle $u$ (while the rest are stationary) leads to edge ratios $\delta_{ut}=M, \delta_{vt}=\delta_{wt}=Id$. However, the result is still locally injective. By the barycentric blending, any line originally parallel to $e_{ij}$ in $t$ is transformed by matrices $M^d$, with varying $d<\frac{1}{2}$, and thus closer to $Id$ than the transformation $M^{\frac{1}{2}}$ of $e_{ij}$. In this case, $e_{ij}$ would be more curved inwards than the other parallel lines within. Thus, in (c), when $M$ is made even more extreme, the target black circular arc from edge $e_{ij}$ and the less-curved blue curve transformed by $M^d$ intersect, causing a loss of injectivity.}
    \label{fig:local-injectivity}
 \end{figure}

\paragraph*{Local injectivity} as mentioned in Sec.~\ref{subsec:mobius-interpolator}, our interpolator is not formally guaranteed to be locally injective. In fact, as we demonstrate in Fig.~\ref{fig:local-injectivity}, this might be the case even if the deformed triangles are not flipped. This happens when the ratios $\delta$ are very different between the edges of the same triangle, which eventually results from a big variation in the \mobius transformation between neighboring triangles. Since parameterization algorithms try to avoid such variations with regularization, we do not expect this to occur often in practice. 

\subsection{Comparisons}
\begin{figure*}
    \centering
    \includegraphics[width=\linewidth]{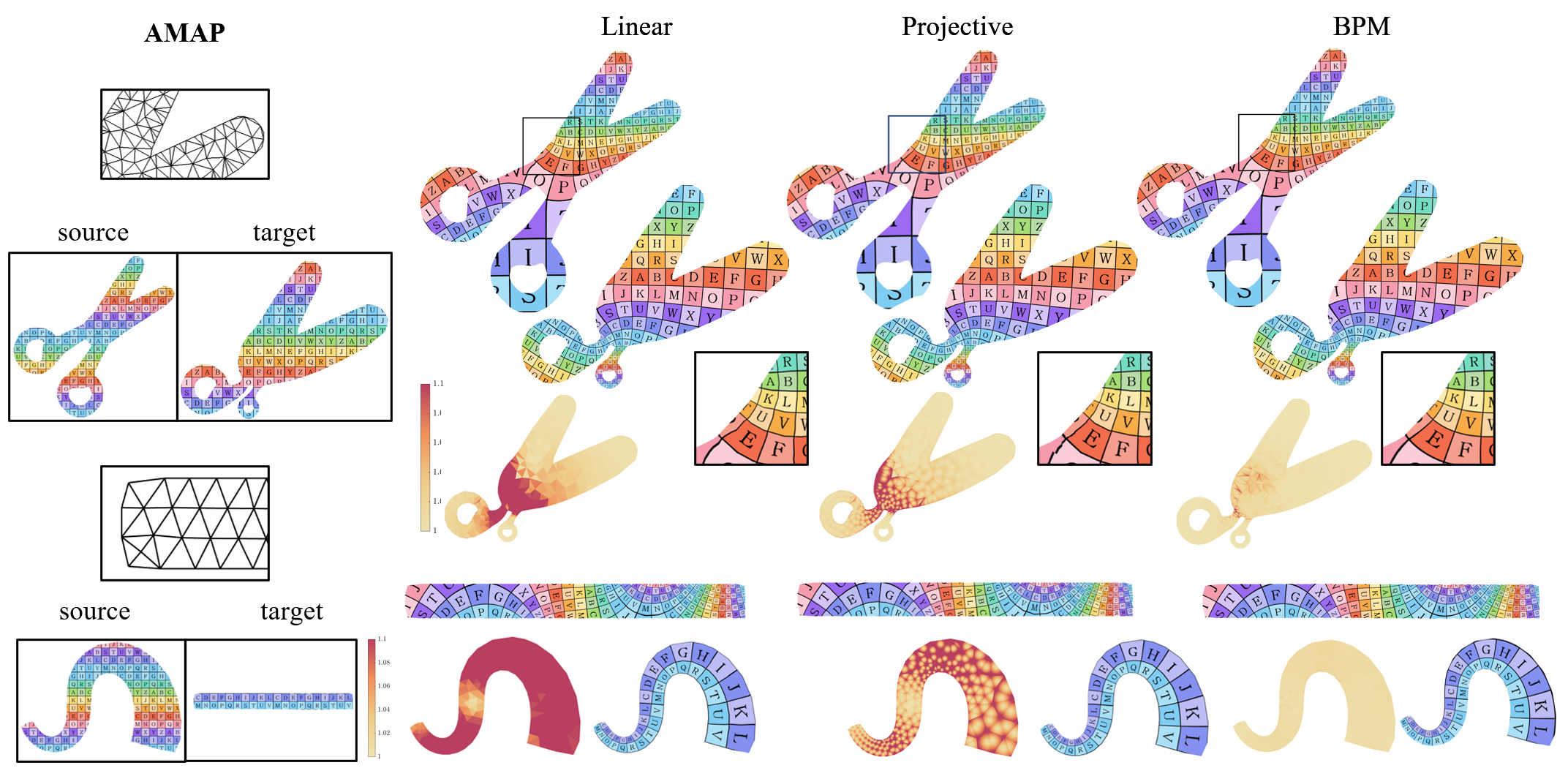}  
    \caption{We compare BPM to linear and projective maps for planar meshes, on input deformations computed using the AMAP method. Note the artefacts in the linear map, and the discontinuities in the projective map, highlighted in the zoomed images. Further, note that our approach yields lower quasi-conformal distortion compared to the alternatives.}
    \label{fig:comparison_amap}
\end{figure*}

\begin{figure*}
    \centering
    \includegraphics[width=\linewidth]{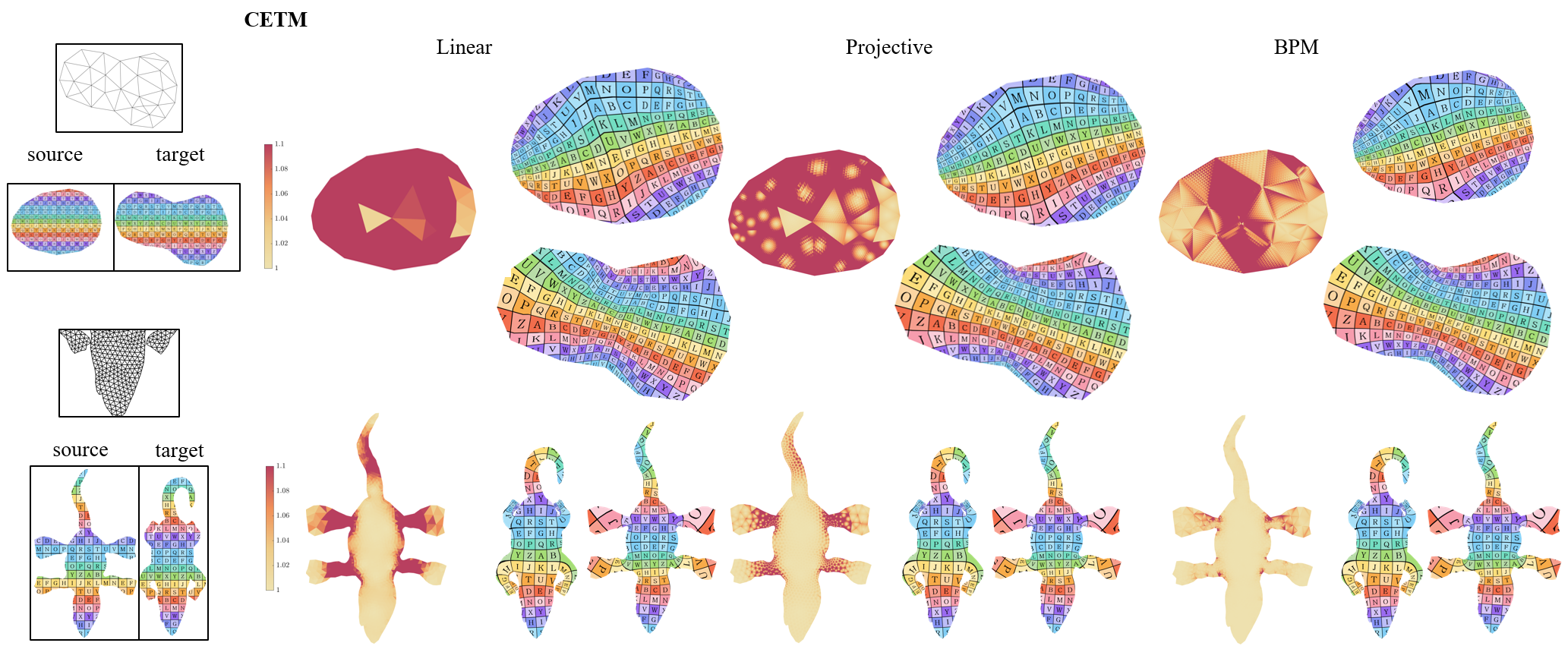}  
     \includegraphics[width=\linewidth]{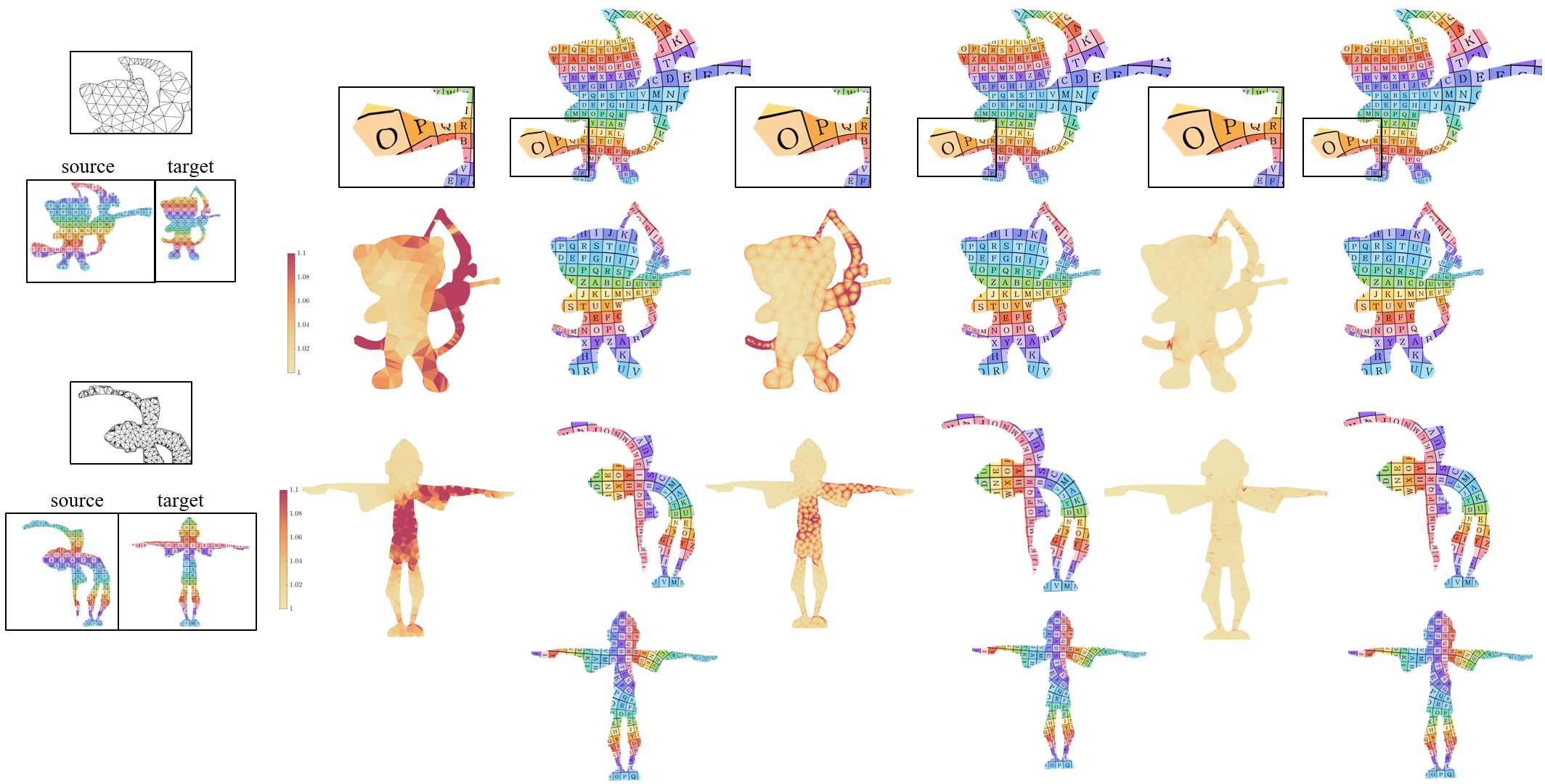}  
    \caption{We compare BPM to linear and projective maps, on two planar shapes which are conformally equivalent (CETM). On this data, our interpolator is comparable to the projective approach, leading to similar texture transfers but lower QC errors.}
    \label{fig:comparison_cetm}
\end{figure*}

\begin{figure*}
    \centering
    \includegraphics[width=1\linewidth]{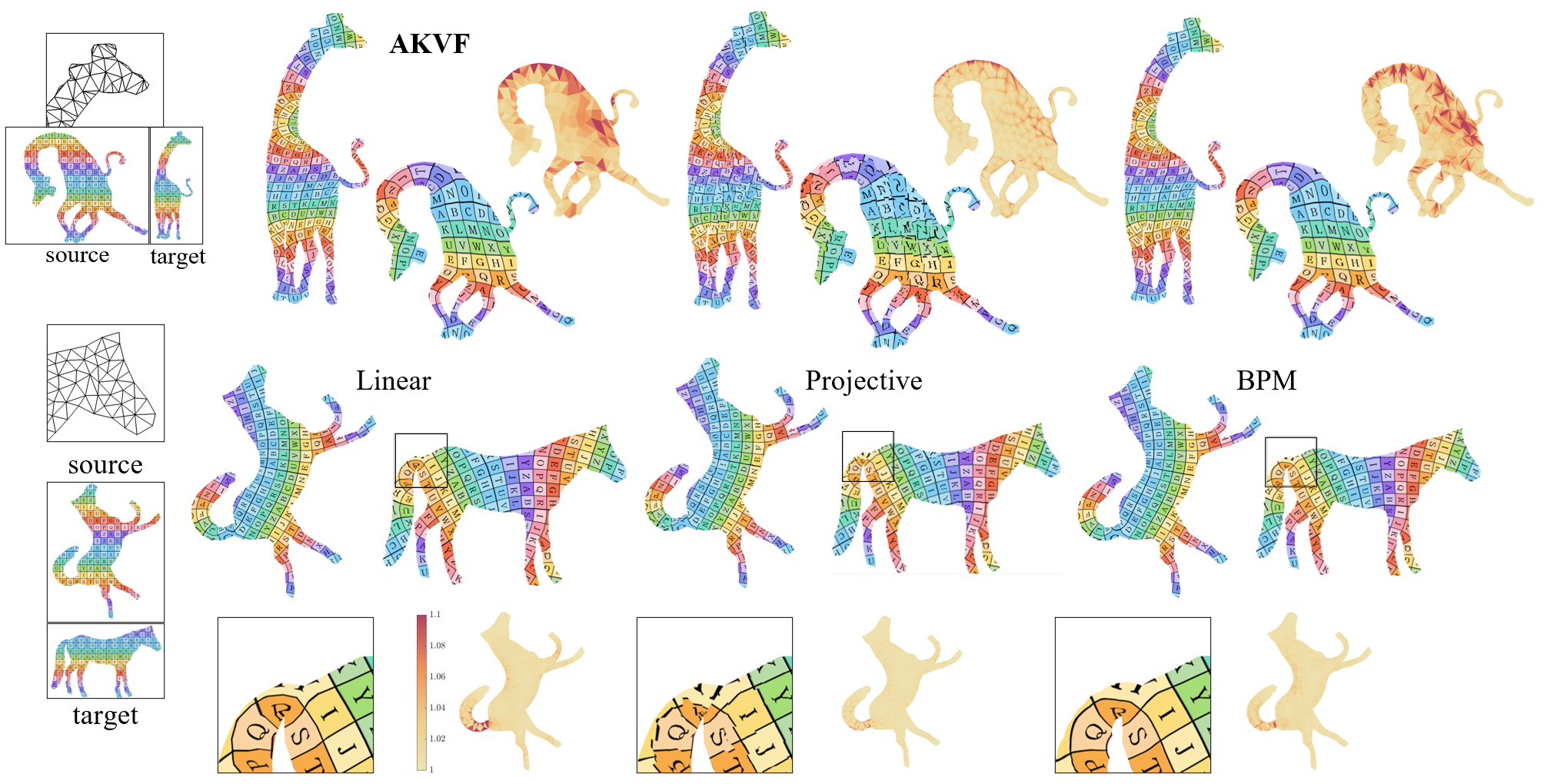}  
    \caption{We compare BPM to linear and projective maps, on planar planar input computed using AKVF. Here, the maps are strongly non-conformal, leading to visible discontinuities in the projective map, whereas our approach leads to smooth results.}
    \label{fig:comparison_kvf}
\end{figure*}

\begin{figure*}
    \centering
    \includegraphics[width=\linewidth]{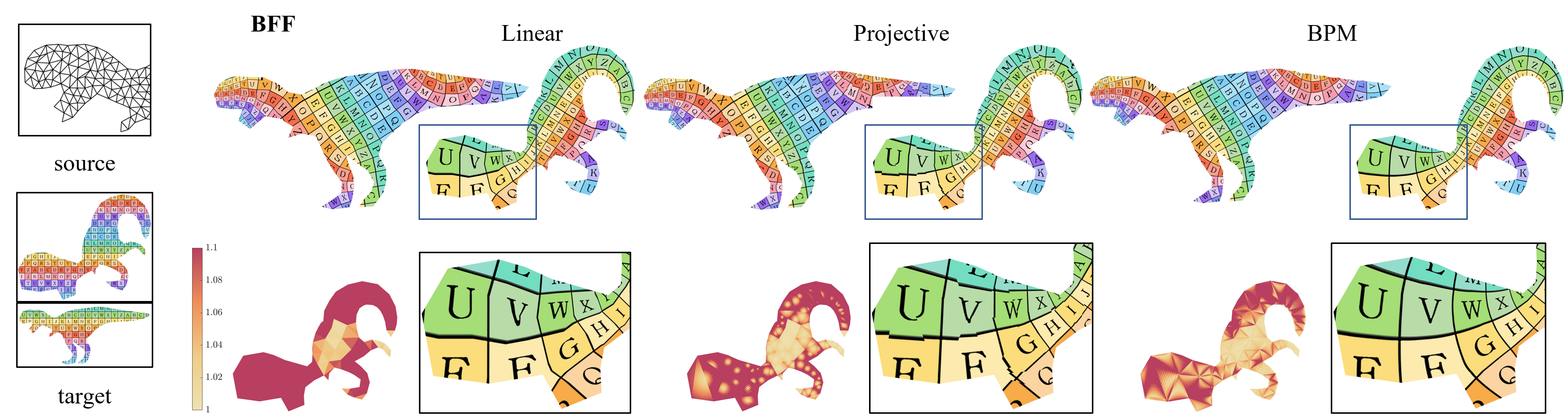}  
    \caption{We compare BPM to linear and projective maps, on input computed using BFF. Here, our approach achieves similar QC distortion as the projective approach. However, since the map is not \emph{exactly} discrete conformal, the projective map leads to discontinuities.}
    \label{fig:comparison_bff}
\end{figure*}

\paragraph*{Interpolators on triangles.}
We compare our approach to PL and projective interpolation, for inputs created with a variety of deformation methods (AMAP, CETM, BFF, AKVF, CC).
\rev{The projective interpolation requires the computation of scaling factors per vertex, which we compute individually per triangle. Note that for meshes that are not CETM, the scaling factors do not agree between different triangles sharing vertices, and therefore the interpolation can be discontinuous.}
We show in Figs.~\ref{fig:comparison_amap},~\ref{fig:comparison_cetm},~\ref{fig:comparison_bff},~\ref{fig:comparison_kvf} the resulting texture maps, as well as the QC distortion for each example. Note that for discrete conformal maps (CETM), and for maps that are close to conformal (BFF, PCM), both the projective interpolation and our approach achieve a good result, though our QC error is lower. Furthermore, our method is applicable to \emph{any} discrete map, whereas projective interpolation is discontinuous for non-CETM maps. This is clearly visible for meshes deformed using AKVF, which can induce significant angle distortion (see Fig.~\ref{fig:comparison_kvf}). Compared to PL interpolation, our map is smoother even for very coarse triangulations (see also Fig.~\ref{fig:multiple-res}).

\begin{figure*}
    \centering
    \includegraphics[width=1.0\linewidth]{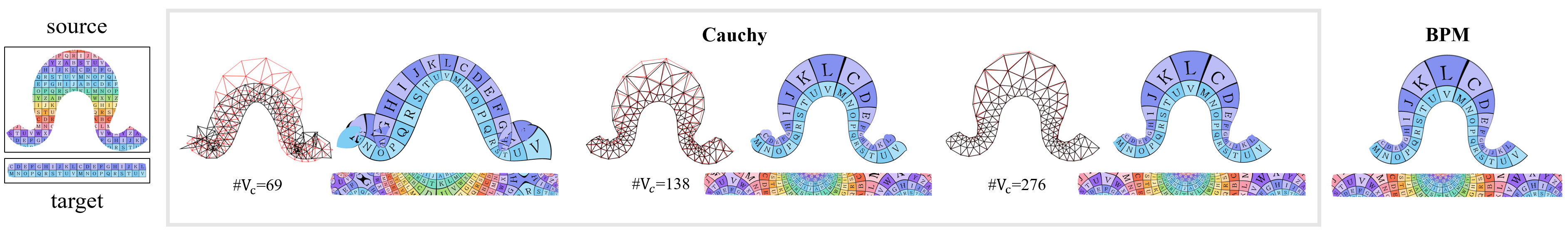}  
    \caption{We compare our approach to computing a smooth map using Cauchy Coordinates, with the input vertex map as constraints and $V_C$ cage vertices. Note that the result highly depends on the number of cage vertices. Using a number that is too small (Cauchy, left), there are not enough degrees of freedom to reproduce the constraints and the mapping becomes a double cover. Using too many cage vertices (Cauchy, center, right) leads to visible oscillations near the boundary. Our approach (right) leads to a smooth interpolation, which is non-oscillatory, does not require additional degrees of freedom, and is closed-form. }
    \label{fig:comparison_cc}
\end{figure*}

\paragraph*{Continuous interpolators.}
Instead of interpolating each triangle separately, or by blending, we attempt to use a continuous interpolator with constraints. Namely, we use a method for which the map is given on the full source triangulation domain (and not only on the vertices), and constrain the vertices to the locations prescribed by the discrete input map. We use Cauchy Coordinates as a smooth interpolator, as it is exactly holomorphic. 
Fig.~\ref{fig:comparison_cc} shows the result of the comparison. On a coarse mesh, if we use a small number of vertices for the cage, the constraints on the vertices cannot be achieved. If, on the other hand, we use a large number of cage vertices, the map generates poles and overlaps. Furthermore, deformation with Cauchy Coordinates is only feasible for a mesh with a small number of vertices, as it is a global approach, that requires solving a linear system with a dense matrix. Hence, our local closed-form approach is a better alternative.


\subsection{Application to texture mapping}
\begin{figure}
    \centering
    \includegraphics[width=1\linewidth]{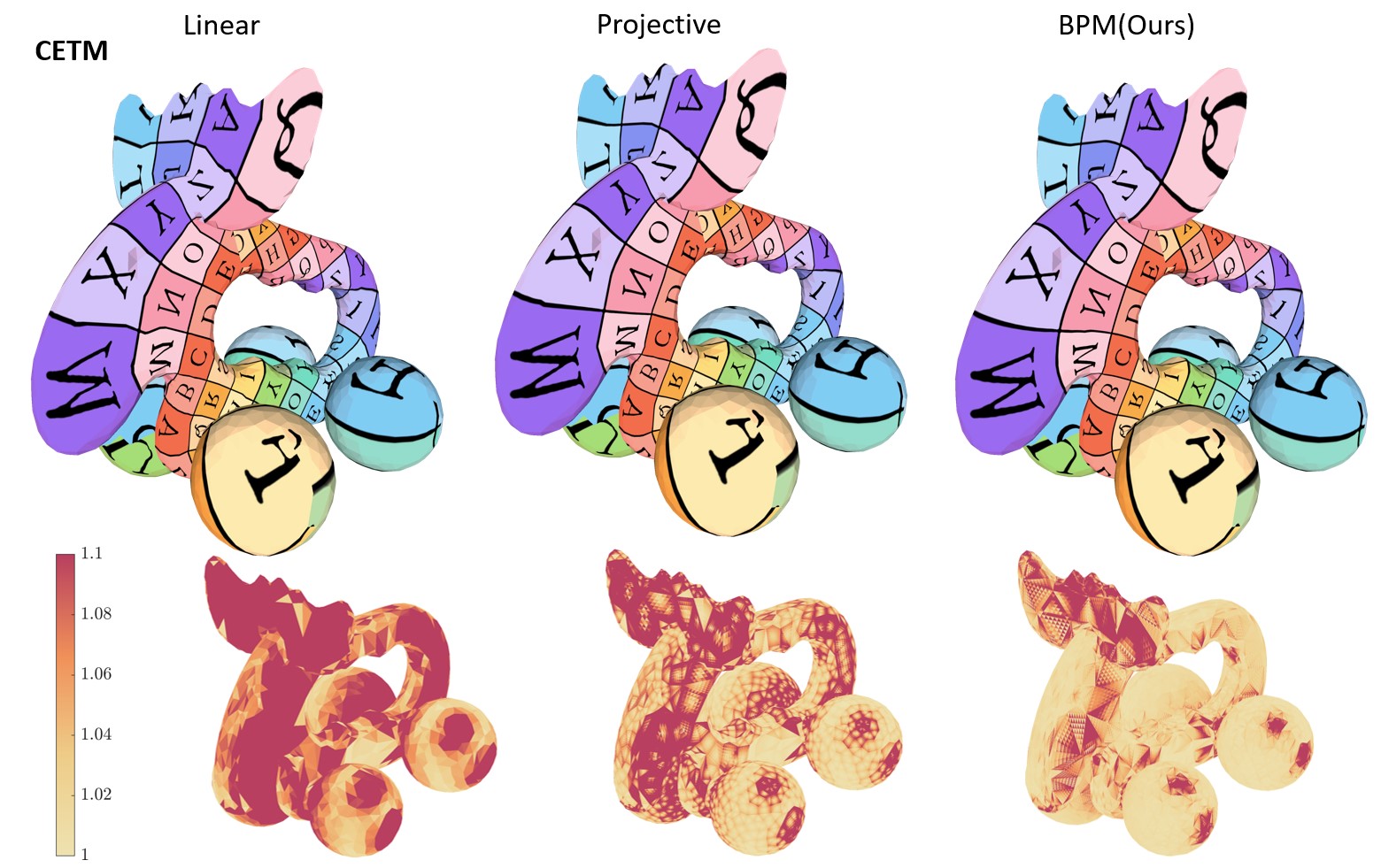}  
    \includegraphics[width=1\linewidth]{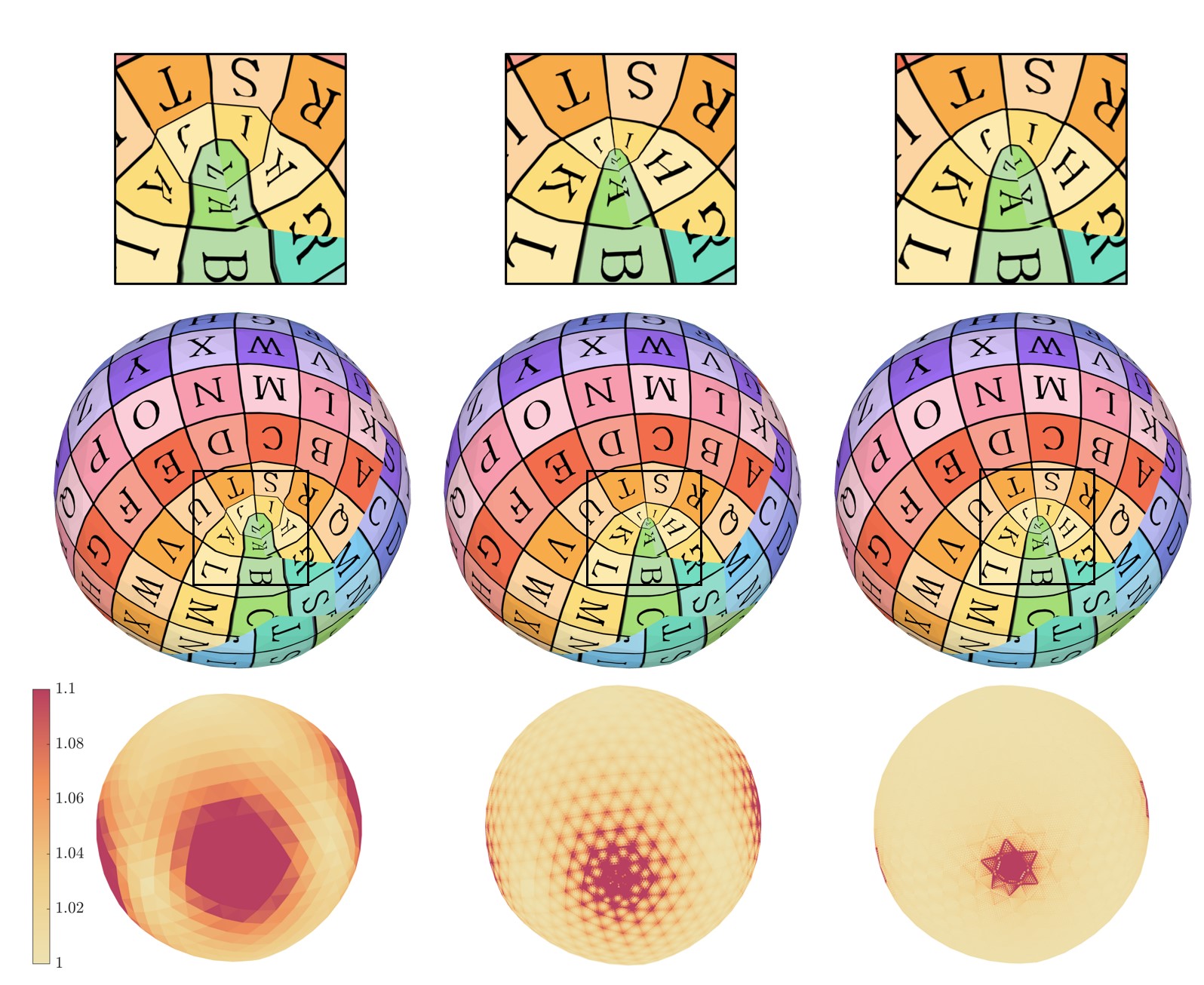}  
    \includegraphics[width=1\linewidth]{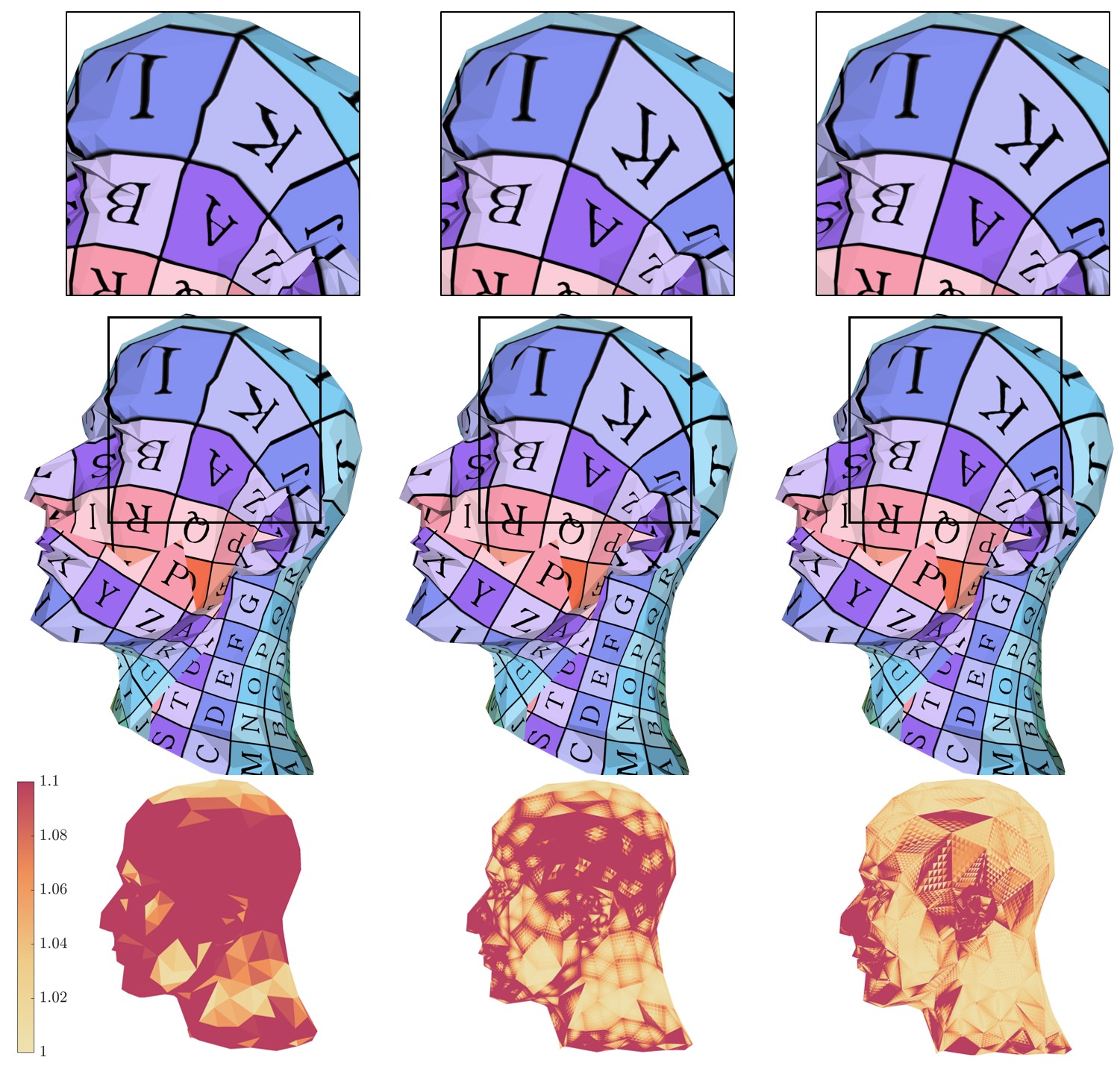}  
    \caption{Applying BPM for interpolating texture coordinates generated using CETM. Projective interpolation is comparable to BPM for CETM inputs, although it generates more artifacts and QC errors.}
    \label{fig:comparison_3d_cetm}
\end{figure}

\begin{figure}
    \centering
    \includegraphics[width=1\linewidth]{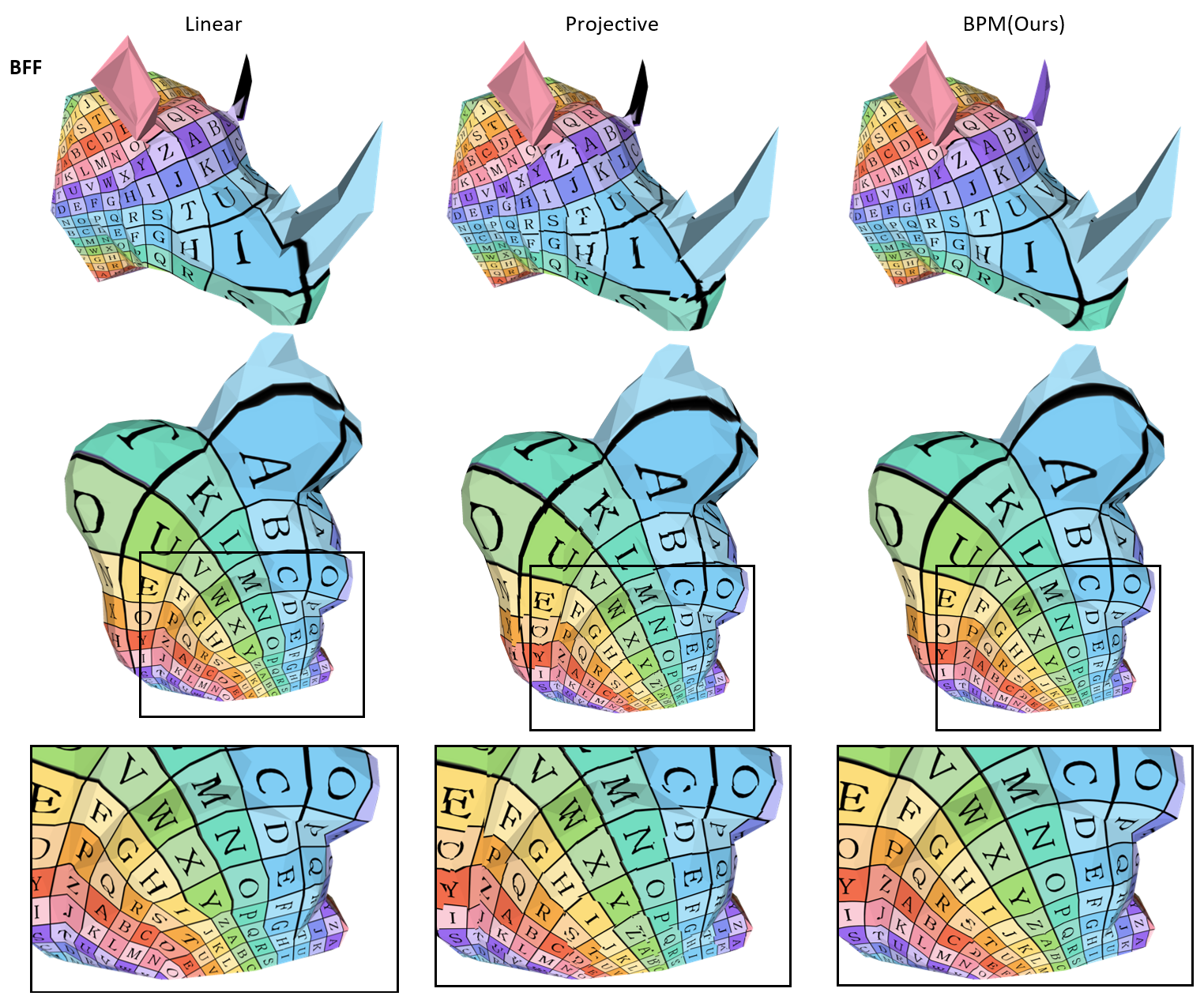}  
    \includegraphics[width=1\linewidth]{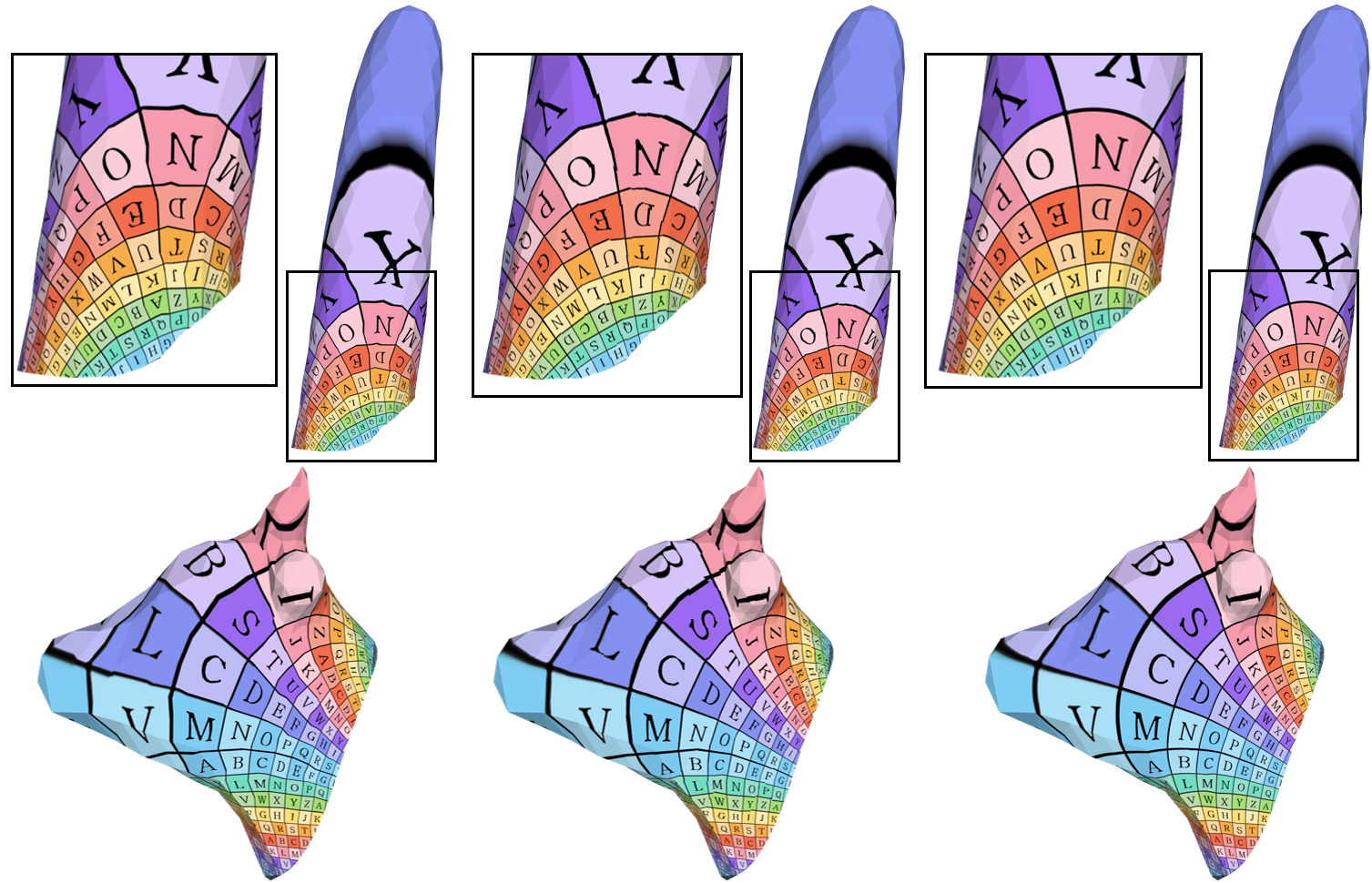}  
    \caption{Applying BPM for interpolating texture coordinates generated using BFF. Note the considerably smoother texture achieved by our approach, compared to piecewise-linear and projective interpolations.}
    \label{fig:comparison_3d_bff}
\end{figure}

\begin{figure}
    \centering
    \includegraphics[width=1\linewidth]{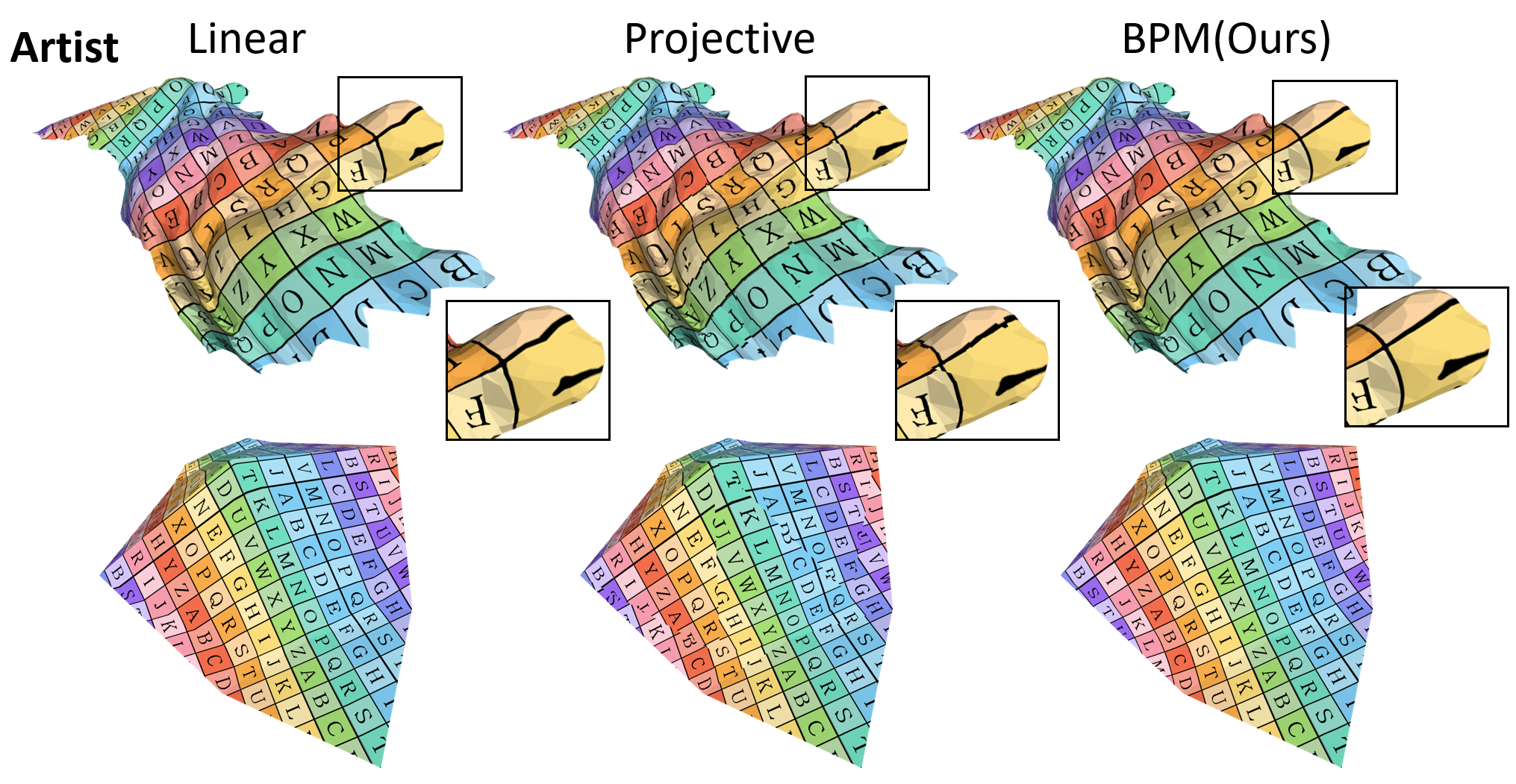}  
   \caption{Interpolation for the artist-UV models from the data-set in ~\cite{shay2022dataset}. The projective interpolation is discontinuous and generates artifacts.}
    \label{fig:Artist_3D}
\end{figure}

Using the intrinsic formulation presented in Sec.~\ref{sec:curved-surfaces} we interpolate the texture coordinates of 3D meshes, leading to considerably smoother textures compared to the alternatives (PL and projective). We demonstrate this in Figs.~\ref{fig:comparison_3d_cetm},~\ref{fig:comparison_3d_bff},~\ref{fig:Artist_3D}, where the inputs are generated using CETM, BFF, and designed by artists, respectively. For CETM, the results are comparable to the projective interpolation, yet our approach achieves lower QC errors, and somewhat smoother outputs. For BFF and artists' generated parameterizations, the projective interpolation is discontinuous, and our results are considerably smoother than both the linear and projective approaches.



\section{Conclusion and Future Work}
We presented a blending scheme (BPM) of \mobius transformations that interpolates a discrete map 
between triangulations to a continuous map on the input domain. Our scheme leads to small quasi-conformal errors when
the input discrete map is close to conformal, and is applicable to \emph{any} discrete input map. We additionally showed that our blending scheme can be done \emph{intrinsically}, thus allowing non-linear interpolation of the texture coordinates of a 3D mesh. 
In the future we plan to explore other applications for our interpolation scheme, such as surface to surface, spherical parameterization, etc. In addition, we plan to investigate \emph{time interpolation} in this setting, as well as generalizing our scheme to blends where the input map is \emph{approximated} instead of interpolated. Finally, we aim to derive theoretical bounds for the QC error of our blends, and classify the conditions under which the map is provably bijective. 

\section{Acknowledgments}
Mirela Ben-Chen acknowledges the support of the Israel Science Foundation (grant No. 1073/21).    

\printbibliography

\clearpage

\appendix

\subsection{Appendix A: Limits of edge barycentric coordinates}
\label{sec:app_a}
The edge weights $B_e(z)$ that we use in Equation~\eqref{eq:ratio-interpolator} are given in terms of the inverse distance to the edge $d(z,e)^{-1}$, which diverges as $z$ approaches the edge. However, the normalized barycentric coordinates: 
\begin{equation}
    \gamma_e(z) = \frac{B_e(z)}{B_{ij}(z)+B_{jk}(z)+B_{ki}(z)}
\end{equation}
have a well-defined limit as $z$ approaches the edge (but not the vertices). \rev{To avoid reaching infinity on the edge}, a simple calculation shows that the coordinates can be computed using only the distances $r_e(z) = d(e,z)$:

\begin{equation}
\begin{aligned}
\label{eq:gammas}
\gamma_{ij}(z) &= \frac{r_{jk}(z) r_{ki}(z) }{s(z)}, \quad \gamma_{jk}(z) = \frac{r_{ij}(z) r_{ki}(z) }{s(z)}, 
\quad \gamma_{ki}(z) = \frac{r_{ij}(z) r_{jk}(z) }{s(z)}, \\
s(z) &= r_{jk}(z)r_{ki}(z) + r_{ij}(z)r_{jk}(z) + r_{ij}(z)r_{ki}(z)    
\end{aligned}
\end{equation}

It is easy to check that if only one of the $r$ quantities goes to 0, i.e., $z$ approaches an edge but not a vertex, the coordinates behave as required, i.e., equal to $1$ on the corresponding edge, and to $0$ on the other two.
However, when $z$ approaches a vertex, the coordinates are still undefined. 
Note that in our scheme the vertices are \emph{interpolated by definition}, and therefore we do not need to use the coordinates to map the original vertices. In practice, we use an epsilon value on the order of machine precision to check if the mapped point corresponds to an input vertex.
We have not encountered any numerical instabilities with this approach.

\subsection{Appendix B: Proof for equation \eqref{eq:log-mobius-ratio}} 
\label{sec:app_b}
We minimize $\lVert \md - I \rVert_F^2$ where $\md \in \{ -\md_{tu},\md_{tu}\}$, \rev{in terms of the Frobenius norm}:
\begin{equation} 
\begin{aligned}
\lVert \md - I \rVert_F^2 = &\text{Tr}((\md - I)^*(\md - I)) = \\
& tr(\md^*\md)-tr(\md^* + \md) + tr(I) & = \\
& tr(\md^*\md)-tr(2\Re(\md)) + tr(I).
\end{aligned}
\end{equation}
This term is minimized for $max(tr(2\Re(\md)))$, and thus $\lmd_{tu} = log\left(\text{Sign}(\text{Tr}(\Re (\md_{tu})))\md_{tu}\right)$.

\rev{\subsection{Appendix C: Pseudo Code}} 
\label{sec:app_c}

We give a pseudo-code description of our interpolator, where Alg.~\ref{alg:bpm} computes the planar-to-planar interpolation, and Alg.~\ref{alg:bpmcurved} computes the curved-surface interpolation (Sec.~\ref{sec:curved-surfaces}).
\begin{algorithm}
  \small
  \caption{ApplyMoebius}
  \SetKwInOut{Input}{inputs}
  \SetKwInOut{Output}{output}
  \SetKwProg{ApplyMoebius}{ApplyMoebius}{}{}

  \ApplyMoebius{$(z,M)$}{
    \Input{ A point $z \tin \xC$, a \mobius matrix M $\tin \xC^{2\times 2}$}
    \Output{$w \in \xC$}
    $\begin{bmatrix} w_1 \\w_2 \end{bmatrix}$ = M$\begin{bmatrix} z \\1 \end{bmatrix}$ \\
    $w = \frac{w_1}{w_2} $
   
    \KwRet{w}\
  }
\end{algorithm}
\begin{algorithm}
  \small
  \caption{BPM}
  \SetKwInOut{Input}{inputs}
  \SetKwInOut{Output}{output}
  \SetKwProg{BPM}{BPM}{}{}

  \BPM{$(F)$}{
    \Input{ A discrete map $\dmap{F}\!:\!\coord{Z}\!\to\!\coord{W}$}
    \Output{a \emph{continuous} map $f:\cmap{Z}\rightarrow\mathbb{C}$}
    Compute the PCM map $M(F)$ (Sec. \ref{sec:Piecewise-Compatible})\\
    \ForEach{$t \tin \mT$}{%
    \ForEach{$z\in t$}{%
    $u,v,w$ = neighboring triangles of $t$\\
      $M_z=MoebiusInterpolator(z,Z,M_t, M_u, M_v, M_w)$\
      $f(z) = ApplyMoebius(z, M_z)$    
    }
    }
    \KwRet{$f$}\
  }
  \label{alg:bpm}
\end{algorithm}

\begin{algorithm}
  \small
  \caption{BPMCurved}
  \SetKwInOut{Input}{inputs}
  \SetKwInOut{Output}{output}
  \SetKwProg{BPMCurved}{BPMCurved}{}{}

  \BPMCurved{$(F)$}{
    \Input{ A discrete map $\dmap{F}\!:\!\coord{X}\!\to\!\coord{W}$}
    \Output{a \emph{continuous} map $f:\cmap{X}\rightarrow\mathbb{C}$}
    \ForEach{$t \tin \mT$}{%
    Compute discrete isometric embedding $\tilde{Z}_t$ (Sec. \ref{sec:curved-surfaces}) \\
    Compute discrete map $\tilde{\dmap{F}_t}:\tilde{\coord{Z}}_t \rightarrow \coord{W}$ \\
    Compute the PCM map $M(\tilde{F}_t)$ (Sec. \ref{sec:Piecewise-Compatible})\\
    \ForEach{$x\in t$}{%
    $z = \tilde{Z}_t(x)$ (embed $x$) \\
    $O(z, M(\tilde{F}_t))=MoebiusInterpolator(z,\tilde{Z}_t,M_t, M_u, M_v, M_w)$\\
    $f(z) = ApplyMoebius(z,O(z, M(\tilde{F}_t)))$
    }}
    \KwRet{$f$}\
  }
  \label{alg:bpmcurved}
\end{algorithm}

\begin{algorithm}
  \small
  \caption{MoebiusInterpolator}
  \SetKwInOut{Input}{inputs}
  \SetKwInOut{Output}{output}
  \SetKwProg{MoebiusInterpolator}{MoebiusInterpolator}{}{}

  \MoebiusInterpolator{$z,Z_t,M_t,M_u,M_v,M_w$}{
    \Input{ A point $z \tin \xC $ ,$Z_t$, The embedding of triangle $t=ijk$ and its neighbors $u,v,w$ (Fig. \ref{fig:notation}), $M_t,M_u,M_v,M_w \tin \xC^{2\times 2}$ the \mobius matrices}
    \Output{$M_z$ the \mobius matrix interpolator at $z$}
    \tcc{\mobius ratios (Eq. \eqref{eq:delta})}
     $\md_{ut} = M_u M_t^{-1}$, $\md_{vt} = M_v M_t^{-1}$, $\md_{wt} = M_w M_t^{-1}$\\
    \tcc{log \mobius ratios (Eq. \eqref{eq:log-mobius-ratio})}
    $ \lmd_{ut} = \log\left(\text{Sign}(\text{Tr}(\Re (\md_{ut}))) \cdot \md_{ut}\right)  $\\
    $ \lmd_{vt} = \log\left(\text{Sign}(\text{Tr}(\Re (\md_{vt}))) \cdot \md_{vt}\right)  $\\
    $ \lmd_{wt} = \log\left(\text{Sign}(\text{Tr}(\Re (\md_{wt}))) \cdot \md_{wt}\right)  $\\ 
    \tcc{Barycentric coord. (Eq.~\eqref{eq:gammas})}
    $s(z) = r_{jk}(z)r_{ki}(z) + r_{ij}(z)r_{jk}(z) + r_{ij}(z)r_{ki}(z) $\
    $\gamma_{ij}(z) = \frac{r_{jk}(z) r_{ki}(z) }{s(z)} $\
    $\gamma_{jk}(z) = \frac{r_{ij}(z) r_{ki}(z) }{s(z)} $\
    $\gamma_{ki}(z) = \frac{r_{ij}(z) r_{jk}(z) }{s(z)} $\\
    \tcc{Blended log ratio (Eq.\eqref{eq:ratio-interpolator})}
    $\lmd_{t}(z,M) = \gamma_{ij}(z)\lmd_{ut}+\gamma_{jk}(z)\lmd_{vt}+\gamma_{ki}(z)\lmd_{wt}$\\
    \tcc{\mobius interpolator (Eq. \eqref{eq:mobius-interpolator})}
    $M_z =\exp\left(\frac{1}{2}\lmd_{t}(z,M)\right)M_t$\\
    \KwRet{$M_z$}
  }
\end{algorithm}

\end{document}